\documentclass[letterpaper,twocolumn,10pt]{article}
\usepackage{usenix-2020-09}

\usepackage{graphicx}
\usepackage{xargs}
\usepackage{xcolor}
\usepackage[colorinlistoftodos,prependcaption,textsize=tiny]{todonotes}
\usepackage{placeins}
\usepackage{csvsimple}
\usepackage{algorithm} 
\usepackage{enumitem}
\usepackage{amsmath}
\usepackage{algpseudocode}
\algnewcommand\TEST{\mathbin{\|}}
\usepackage{underscore}
\usepackage{xcolor,colortbl}
\usepackage{listings}
\usepackage{booktabs}                             
\usepackage{multirow}
\usepackage{url}
\usepackage{tikz}
\usepackage{msc}
\usepackage{adjustbox}
\usetikzlibrary{positioning, arrows.meta, matrix, backgrounds, shapes.misc}
\usetikzlibrary{arrows}
\usepackage[labelfont=bf]{caption}
\makeatother

\algdef{SE}[VARIABLES]{Variables}{EndVariables}
   {\algorithmicvariables}
   {\algorithmicend\ \algorithmicvariables}
\algnewcommand{\algorithmicvariables}{\textbf{global variables}}

\makeatletter
\g@addto@macro{\UrlBreaks}{\UrlOrds}
\makeatother

\begin{document}

\date{}

\definecolor{blue(pigment)}{rgb}{0.2, 0.2, 0.6}
\definecolor{green(pigment)}{rgb}{0.0, 0.65, 0.31}
\definecolor{red(pigment)}{rgb}{0.93, 0.11, 0.14}
\lstdefinestyle{base}{
  language=C,
  emptylines=1,
  numbers=left,
  firstnumber=1,
  stepnumber=1,
  escapeinside={\%}{\%},
  columns=fullflexible,
  breaklines=true,
  basicstyle=\small\ttfamily,
  moredelim=**[is][\color{red(pigment)}]{@}{@},
  moredelim=**[is][\color{blue(pigment)}]{`}{`},
  moredelim=**[is][\color{green(pigment)}]{!}{!}
}

\title{\Large \bf Security and Privacy Analysis of Samsung's Crowd-Sourced Bluetooth Location Tracking System}


\author{
{\rm Tingfeng Yu} \thanks{Artefact available on GitHub: \url{https://github.com/pogen300/galaxy_smarttag_research}}
\\
 School of Computing\\
The Australian National University
 \and
 {\rm James Henderson}\\
 School of Computing\\
 The Australian National University
 \and 
 {\rm Alwen Tiu}\\
 School of Computing\\
 The Australian National University
 \and 
 {\rm Thomas Haines}\\
 School of Computing\\
 The Australian National University
 } 

\maketitle

\begin{abstract}
We present a detailed analysis of Samsung's Offline Finding (OF) protocol, which is part of Samsung's Find My Mobile system for locating Samsung mobile devices and Galaxy SmartTags.  The OF protocol uses Bluetooth Low Energy (BLE) to broadcast a unique beacon for a lost device. This beacon is then picked up by nearby Samsung phones or tablets (the {\em helper} devices), which then forward the beacon and the location it was detected at, to a vendor server. The owner of a lost device can then query the server to locate their device. 
We examine several security and privacy related properties of the OF protocol and its implementation.
These include: the feasibility of tracking an OF device through its BLE data, the feasibility of unwanted tracking of a person by exploiting the OF network, the feasibility for the vendor to de-anonymise location reports to determine the locations of the owner or the helper devices, and the feasibility for an attacker to compromise the integrity of the location reports. Our findings suggest that there are privacy risks on all accounts, arising from issues in the design and the implementation of the OF protocol.
\end{abstract}

\section{Introduction}
\label{sec:introduction}

Portable devices such as smart phones and tablets often come with a feature that allows their owner to find those devices when they are lost, typically through the use of a web portal provided by their vendors, 
such as Google's Find My Device \cite{googlefindmy}, Samsung's Find My Mobile (FMM) \cite{samsungfindmy} and Apple's Find My \cite{applefindmy}. A typical requirement for such a feature to work is that the lost device must be connected to the internet so that it can send its location report to a vendor server in the event that its owner flags the device as lost. 
In recent years, mobile device manufacturers such as Samsung and Apple have extended their lost-device tracking systems with an {\em offline finding} (OF) feature, which allows a lost mobile device to be found even when it does not have an internet connection. Both Apple and Samsung OF features share two key elements: the use of Bluetooth Low Energy (BLE) for short range transmission of data between devices of a vendor, and crucially, an extensive network of (internet-connected) mobile devices (which we call {\em helper devices}) that relay location information to a vendor controlled server. We refer to the latter as the {\em OF network}. The basic idea is quite simple: when a lost device loses its internet connection, it starts broadcasting a unique beacon over BLE, which is then picked up by nearby helper devices participating in the OF network, who then forward the beacon and the location it is found to a vendor server. 
In this work, we are mainly concerned with Samsung's FMM Offline Finding (OF) feature \cite{samsungfmmapp}, which was introduced in 2020. An owner can track their devices' locations through Samsung FMM application running in a Samsung mobile device (e.g., a phone or a tablet).

In 2021, Samsung released the Galaxy SmartTag \cite{wiki:smarttag}, which is a small BLE tracker that can be attached to various items to keep track of their locations.
SmartTags are not capable of connecting to the Internet, so they rely on the OF network for long range location tracking (outside the range of BLE). SmartTags are registered and controlled through SmartThings, which is an umbrella control and management platform for a large variety of smart devices and home appliances. OF is also supported for SmartTags using the "SmartThings Find" add-on which works in conjunction with FMM. 

Devices in Samsung's OF network can be categorized into three roles: the owner device, the helper device, and the lost device. A mobile device can be registered to the Samsung OF network through the FMM app, while a SmartTag can be registered through Samsung SmartThings app. Each registered device is linked to the owner's account under which it was registered from. 
When a registered device loses internet connectivity, or in the case of SmartTags, when it is out of the BLE range from its owner device, it broadcasts certain data over BLE periodically. 
This data contains a rotating identifier, called the {\em privacy ID}, which is unique to the lost device and which, in theory, can only be linked to its owner by Samsung and the owner device. 
The helper devices consist of both Samsung devices (phones and tablets), and some third-party devices that support Samsung OF protocol. An active helper device periodically scans for BLE advertisements from nearby OF devices and reports their locations to a location server. The location reports of the lost devices will be downloaded onto the owner device when the owner queries their locations. 
The effectiveness of the OF feature depends on the size of its OF network, 
which in the case of Samsung OF, is estimated to have around 200 million active helper devices~\cite{Samsung2022} in 2022. 

In the following, we shall refer to end-user devices, such as mobile phones and SmartTags, that participate in Samsung OF network as OF devices. 
Our work was driven by the following research questions: 
\begin{description}
    \item[(RQ1)] {\em Identification of an OF device.} Can an OF device be identified through its BLE data? 
    \item[(RQ2)] {\em Unwanted tracking.} Can Samsung OF network be misused for unwanted tracking of a person or an object by a party other than Samsung? 
    \item[(RQ3)] {\em End-to-end location privacy.} To what extent does the design of the OF protocol protect the location privacy of the lost devices and the helper devices from the service provider (Samsung)? 
    \item[(RQ4)] {\em Location report integrity.} Is it possible for an actor (other than the owner and the vendor) to forge a location report of a lost device? 
\end{description}
RQ1 concerns the privacy protection of the owner of an OF device against (long term or short term) location tracking through the BLE data emitted by the device by an adversary.
RQ2 addresses an attack scenario where 
a tag is used by its owner to stalk a person without their consent~\cite{airtagstalking}. 
RQ3 raises the question as to what extent Samsung is aware of the movement of both the owner devices and the helper devices. 
RQ4 is more of a security (integrity) issue rather than a privacy one and addresses a scenario where the attacker intentionally disrupts the location tracking capability of the OF network through false location reports. 

Our main contributions are as follows: 
\begin{itemize}
    \item We provide the first comprehensive reverse-engineering of Samsung OF protocol
    that allows us to answer definitively the research questions (RQ1 to RQ4) raised above. 
    
    \item We identified several vulnerabilities that would allow an attacker to link BLE packets observed from a target device over multiple observations, through BLE interactions only, allowing a long term identification of an OF device through its BLE data (RQ1). 
    
    \item Through our analysis of the OF protocol, we managed to impersonate completely a SmartTag to the OF network. This opens the possibility of creating a custom tracking device that can be tuned to circumvent potential anti-tracking mechanisms by the vendor. 
    
    \item Our analysis also confirmed that unwanted tracking (RQ2) is possible. 
    
    \item Our analysis suggests that the vendor does indeed possess the information needed to link an account to a location report (RQ3).  
    Moreover, the vendor server does not appear to check the integrity of the location reports, opening it to manipulation by third parties (RQ4).

\end{itemize}

\emph{\bf Coordinated disclosure.} 
We have reported all the issues we found (see \S\ref{sec:analysis}) to Samsung and followed an industry-standard of a minimum of 90-day embargo period, prior to the publication of these issues. One of the issues we raised concerns the small pool of privacy IDs being used for FMM BLE packets, which has now been assigned a CVE (CVE-2022-33707, SVE-2022-0126). 
The more critical security issues that affect the server-side of Samsung OF network, related to SmartTag registration and the location reporting, have been fixed. Section~\ref{sec:bug-fixes} provides an update on the status of these issues.
Not all the issues we reported have been patched, and in one case, concerning the issue of de-anonymisation through BLE DFU mode (see Section~\ref{ssec:dfu}), the vendor confirmed that they have no plans to issue a fix. However, we believe that there are significant public interests in the disclosure of these issues, both from the end user perspective, so that they can make informed decisions on the use of affected devices, and from a scientific perspective, as some of the issues we discovered arise from design choices that are not specific to a particular vendor.

\emph{\bf Related work.}
The closest to our work is the security and privacy analysis of Apple's FindMy network~\cite{Heinrich21}. Their study uncovered design and implementation flaws that could lead to location correlation attacks and unauthorized access to location histories. They reverse-engineered FindMy protocols and showed that one could create custom tracking devices leveraging on the FindMy network\cite{seemooopenhaystack}.

Several vulnerabilities~\cite{fmmx1} have been identified in an earlier version of Samsung FMM app, that allow, among others, a malicious app installed in the device to manipulate the URL endpoint accessed by the FMM app and to access unprotected broadcast receivers in the FMM app. This analysis was done prior to the introduction of the offline finding features to FMM, so it did not cover the OF related vulnerabilities.  

Both Apple AirTags and Samsung SmartTags use the nRF52 series System on Chips (SoC),
which are known to be vulnerable to a power glitching attack. The firmware of both AirTags and SmartTags have been extracted by researchers using this attack~\cite{Roth22,Samsung-SmartTag-Hack}. While AirTag firmware has been studied extensively~\cite{Roth22}, we are not aware of any reverse-engineering attempt on the SmartTag firmware prior to our work. 


There have been a variety of BLE trackers prior to the introduction of AirTags and SmartTags, such as the Tile tracker. 
Weller et. al.~\cite{Weller20} provides a detailed security and privacy analysis of the security and privacy aspects of these trackers, 
focusing on their interactions with their associated mobile apps and the backend cloud servers for crowd-sourced location tracking. However, they did not analyze the privacy issues arising from the BLE protocols used in these trackers.  

Apple's FindMy network consists of hundreds of millions of active devices, which has raised privacy concerns on whether the network can be abused for malicious tracking. Apple has implemented an anti-tracking framework to detect unknown FindMy trackers for iOS devices. However, Mayberry et al. \cite{Mayberry21} discovered that Apple's tracking detection mechanism can be defeated through a number of techniques. One technique exploits a ``blindspot'' in Apple's anti-tracking algorithm that ignores lost mobile devices, focusing only on detecting lost AirTags. As we shall see later, Samsung's anti-tracking feature suffers from a similar oversight. 

AirGuard is an anti-tracking application developed by SEEMOO lab to protect Android users from BLE trackers that leverage Apple's FindMy network \cite{airguard}. It has a higher success rate and lower false positive rate in detecting and reporting trackers compared to Apple's built-in anti-tracking. 

    A BLE device can be identified by its MAC address. To avoid long-term tracking of a BLE device, vendors often implement anti-tracking mechanisms, such as by randomizing its MAC address. 
    However, issues have been found in implementations of the communication protocols of BLE devices, which can be used to defeat such anti-tracking mechanisms\cite{Celosia19,Celosia19b,Celosia20,Martin19}. 
    Our own investigations show that many of these issues also affect SmartTags.

Another closely related research area is BLE-based contact-tracing applications, 
such as Google-Apple Exposure Notification (GAEN) framework ~\cite{GAEN}, DP3T~\cite{TroncosoBBCCGHJ22}, Singapore's TraceTogether and Australia's COVIDSafe~\cite{Tiu20}. 
The BLE-related attack surface of these applications, i.e., in relation to RQ1, is similar to our findings, e.g., attacks discussed in~\cite{BaumgartnerDFGH20,IovinoVV20,Tiu20,BrightonKnight21} show that devices running these applications can be linked through their BLE data. Another similarity is the ``wormhole'' relay attack demonstrated in~\cite{BaumgartnerDFGH20}, which is similar to the relay attack discussed in Section~\ref{ssec:network-based}. There is, however, a crucial difference between an offline location tracking system and a contact-tracing system. In the latter, the ``location'' information (which takes the form of a list of anonymized user IDs detected in proximity, without any geolocation information) is not sent to users of the applications, but only to a designated health authority. This essentially makes the RQ2, i.e., the possibility of unwanted tracking irrelevant.

\emph{\bf Outline.} The remainder of the paper is structured as follows: Section \ref{sec:background} gives a brief overview of relevant cryptographic and BLE related concepts. 
Section \ref{sec:smarttags} presents technical details of the OF protocol that result from our investigations. This section covers the OF operations for SmartTags. The OF operations for FMM devices are a much simplified version of the SmartTag protocol, so its security and privacy issues are subsumed to those of SmartTags. The interested reader can consult Appendix~\ref{sec:fmm} for details of the FMM protocol. 
In Section \ref{sec:analysis}, we perform a security and privacy analysis on Samsung OF protocol based on our findings discussed in previous sections. In Section~\ref{sec:discussions} we discuss the broader implication of our findings on the design of offline finding protocols in general. Section \ref{sec:conclusion} concludes the paper. For clarity, our presentation of the OF protocol abstracts away some concrete details such the data format and the BLE interface used, but these details are available in the appendices.

\section{Background and methodology}
\label{sec:background}
This section gives a very brief overview of the relevant cryptographic functions used in the Samsung OF protocols and some basic concepts related to BLE. 

\emph{\bf ECDH key exchange and AES block cipher.} 
There are two main cryptographic constructions used in the OF protocol: the Elliptic-curve Diffie-Hellman (ECDH) key exchange protocol and the AES block cipher and its associated encryption modes. 
We explain briefly each of these constructions. For further details, we refer interested readers to \cite{Hankerson2010} for ECDH and \cite{DaemenR02} for the AES algorithm. 

The ECDH builds on the Diffie-Hellman key exchange protocol~\cite{DiffieH76}, where the underlying group operations are defined over an elliptic curve (EC).
Samsung's OF implementation of ECDH uses the elliptic curve Curve25519~\cite{Bernstein06}, which was designed to achieve high speeds at computation without compromising the security strength. The Advanced Encryption Standard (AES) algorithm~\cite{DaemenR02} is a symmetric block cipher that is widely used for data encryption, and as a building block for other cryptographic functions. 
Samsung's FMM and SmartTags implement AES CBC mode cipher with PKCS\#7 padding scheme~\cite{pkcs7} to encrypt/decrypt data for various OF related operations. 

\emph{\bf Bluetooth Low Energy (BLE)}
SmartTags uses BLE~\cite{bluetoothcore}, which is a short-range wireless communication technology, for data transmission. 
The protocol stack of BLE consists of various layers and profiles, of which, the most relevant ones to this work are the Generic Access Profile (GAP) and the Generic Attribute Profile (GATT).
GAP defines the procedures for device discovery and connection establishment. A BLE device can operate in one or more of the following roles:
\begin{itemize}
    \item Advertiser: a device that sends out BLE data that is available to any nearby Bluetooth capable devices.
    \item Observer: a device that listens to BLE advertisement data and may process the data from advertisers.
    \item Central: a device that initiates a connection after receiving advertisement data from an advertiser.
    \item Peripheral: a device that accepts the incoming connection from a central.
\end{itemize}
GATT defines the data organization and exchange over connections between BLE devices. GATT uses a hierarchical structure to organize data. 
A GATT profile may contain multiple services, each contains one or more characteristics. Each characteristic is a container of user data. A characteristic can be followed by descriptors, which provide additional metadata of the characteristic and its value.

BLE has two ways of transferring data: advertising over BLE and data exchange over connections.
Advertising is the process of a BLE device sending out data packets in one-way, while communication over connections allows bidirectional data transfer between the peripheral and the central. Data packets are exchanged through characteristics in the GATT server of the peripheral device.
A BLE device is addressed through its MAC address, which is a 48-bit identifier. There are four types of MAC addresses: Public Address, Random Static Address, Random Private Non-Resolvable Address, and Random Private Resolvable Address. 
A Public Address is registered with IEEE and never changes. A Random Static Address is not registered and remains constant during device runtime. Each Bluetooth-capable device has an {\em Identity Address}, which is either a Public Address or Random Static Address. 
The two types of Random Private Addresses (Non-Resolvable, Resolvable) are used for privacy protection purposes. Random Private Non-Resolvable Addresses are generated completely randomly, whereas Random Private Resolvable Addresses (RPAs) are generated using a key-hashed function from a random seed value and a 16-byte key called the Identity Resolving Key (IRK). The possession of the IRK of a device would also allow one to de-anonymize its RPAs. 

Pairing is the process by which two BLE devices exchange necessary information so that an encrypted connection can be established. BLE has several pairing modes, which are determined by the authentication requirements and input/output (IO) capabilities of the pairing devices. As part of pairing, the IRKs are exchanged, so the devices can identify their respectives RPAs using the IRKs. 
BLE supports different pairing methods to authenticate participants in the pairing procedure. The simplest pairing method, called {\em Just Works}, does not check the authenticitiy of the participants, and has been exploited to steal the IRK of 
a target device stealthily~\cite{Xu19NDSS,Zhang20Usenix,Tiu20}.
 
\emph{\bf Methodology.}
A combination of investigation approaches were used to understand the OF protocol for SmartTags and the FMM app. Devices involved in the analysis include a research laptop equipped with a BLE 4.2 adapter running Ubuntu 20.04 LTS, a number of Samsung mobile phones running Android versions 8.0 - 12,  
(Galaxy S7 Edge running Android 8.0, Galaxy A12 running Android 11, Galaxy S20 and Galaxy S21 Ultra, both running Android 12) 
and SmartTags with firmware version 1.01.26 and 1.02.06. 
We used a combination of application and firmware reverse engineering to study the innerworking of various protocols involved in the OF network, analysis of various logs produced by Android systems and applications, and analysis of both BLE and network traffic between devices and vendor servers. The methodology is quite standard for vulnerability research so we omit the details here.


%
\begin{figure}[!ht]
\resizebox{\columnwidth}{!}{
  \setmsckeyword{}
  \drawframe{no}
  \setlength{\topheaddist}{1pt}
  \setlength{\bottomfootdist}{1pt}
  \begin{msc}[instance distance=1.0cm, label distance=0.2ex]{}
    \declinst{o}{}{Owner}
    \declinst{t}{}{Tag ($pub_{tag},priv_{tag}$)}
    \declinst{s}{}{Server ($pub_{tag}$)}
    \declinst{h}{}{Helper}
    \inlinestart[left inline overlap=1.8cm]{reg}{\parbox{1.7cm}{Registration (\S\ref{sec:smarttag-registration})}}{o}{s} 
    \nextlevel[2]
    \mess{querying tag data}{o}{t}
    \mess[label position=below]{}{t}{o}   
    \nextlevel
    \mess[label position=above right]{key establishment}{o}{s}
    \mess[label position=below]{}{s}{o}
    \nextlevel[2]
    \mess{authentication \&}{o}{t}
    \mess[label position=below]{confirm physical possession}{t}{o}
    \nextlevel[2]
    \mess[label position=above right]{ownership check \&}{o}{s}
    \mess[label position=below right]{finalize registration}{s}{o} 
    \nextlevel 
    \mess{}{o}{t}
    \mess[label position=below]{finalize registration}{t}{o}  
    \nextlevel 
    \inlineend{reg}
    \nextlevel 
    \inlinestart[left inline overlap=1.8cm]{con}{\parbox{1.7cm}{Connected mode (\S\ref{sec:gatt-commands})}}{o}{t} 
    \nextlevel[2]
    \mess{authentication \&}{o}{t}
    \mess[label position=below]{commands}{t}{o}
    \nextlevel 
    \inlineend{con}
    \nextlevel 
    \inlinestart[left inline overlap=1.8cm]{lost}{\parbox{1.7cm}{Lost mode (\S\ref{ssec:lost-and-found})}}{o}{h} 
    \nextlevel[2]
    \mess[label position=above left]{privacy id broadcast}{t}{h}
    \nextlevel
    \mess{location report}{h}{s}
    \nextlevel
    \mess[label position=above left]{query tag location}{o}{s}
    \mess[label position=below]{}{s}{o}
    \nextlevel 
    \inlinestart[left inline overlap=1.8cm]{stalk}{\parbox{1.7cm}{Overmature Lost mode (\S\ref{ssec:unknown-tag})}}{t}{h} 
    \nextlevel[2]
    \mess[label position=above left]{privacy id broadcast}{t}{h} 
    \nextlevel
    \mess{unknown tag check \&}{h}{s}
    \mess[label position=below]{request temp. key}{s}{h}
    \nextlevel[2]
    \mess[label position=above left]{authentication (temp. key) \&}{h}{t}
    \mess[label position=below left]{play sound}{t}{h}
    \nextlevel 
    \inlineend{stalk}
    \inlineend{lost}
  \end{msc}
}
\caption{An overview of Samsung OF protocol for SmartTags} 
\label{fig:subprotocols}
\end{figure}

\section{The offline finding protocol}\label{sec:smarttags}

This section discusses the key findings on the Samsung offline finding (OF) protocol. Here we discuss only the OF protocol for Galaxy SmartTags, but details of the OF protocol for FMM, which is a simplified version of the SmartTag protocol, are available in Appendix~\ref{sec:fmm}.  

Our analysis was performed on SmartTags with firmware versions 1.01.26 and 1.02.06, and the FMM app versions prior to version 7.2.24.12. 


The OF protocol for SmartTags involves both online interactions (over the Internet) with various vendor servers and offline interactions (over BLE) with nearby tags and mobile devices. We discuss here four important subprotocols, which are summarised in Figure~\ref{fig:subprotocols}. The communication between the devices and the server is done through HTTPS, which we assume to be secure. 
There are four principals involved: the owner device, the tag, the vendor server and the helper device.
The subprotocols are as follows:
\begin{enumerate}
\item The registration protocol (\S\ref{sec:smarttag-registration}). This protocol involves the owner device, the tag and the server. The owner initiates the protocol by acquiring relevant tag data, such as the serial number, firmware version, etc., and initiates a key establishment protocol with the server, to derive various symmetric keys that will be used in subsequent interactions with the tag. 

\item The protocol for tags in the {\em connected mode} (\S\ref{sec:gatt-commands}). This protocol is executed right after the registration, and whenever the tag is in the proximity of the owner device after having been out of the BLE range. It contains subprotocols for authenticating the owner device to the tag and vice versa. Once authenticated, various commands and data may be exchanged over BLE.

\item The protocol for tags in the {\em lost mode} (\S\ref{ssec:lost-and-found}). This protocol is triggered after the tag has lost its BLE connection to its owner for less than 24 hours. In this state, it broadcasts anonymized rotating {\em privacy IDs} that trigger nearby helper devices (who are scanning for privacy IDs periodically) to record and report the locations of the recorded privacy IDs to the server. 

\item The protocol for tags in the {\em overmature lost mode} (\S\ref{ssec:unknown-tag}). A tag transitions to the {\em overmature lost mode} after being lost for over 24 hour.
A helper device that detected a tag in the overmature lost mode would initiate an anti-stalking detection process. 
This process aims to enable the helper device to locate a (potentially) tracking tag by playing sound on the tag. To play sound, the helper would need to obtain a temporary key from the server to authenticate itself to the tag. 

\end{enumerate}

\subsection{SmartTag registration}\label{sec:smarttag-registration}
The SmartTag registration protocol requires interacting with an unregistered tag and the vendor server. For the latter, there are actually multiple servers involved, providing services such as user authentication, application related services such as remote attestation, and services related to storing and retrieving location reports. For simplicity, we shall refer to these servers collectively as the vendor server (or simply "the server") in the following discussion. 

The interaction between the owner device and the tag is done through BLE using BLE advertisement and a GATT profile. 
A SmartTag uses two UUIDs to advertise its presence over BLE: FD59 for  non-registered tags, and FD5A for registered tags. 
SmartTags do not support internet connectivity and thus rely on the owner device (typically a mobile phone) to perform various setups over BLE connections. This is done through its GATT profile, which defines various services and characteristics that the tag and its owner device use to exchange data and commands. 
In the following, we shall omit the concrete UUID used for each characteristic in the GATT profile, and refer to it using a symbolic name instead. But detailed characteristic UUIDs can be found in Appendix~\ref{sec:smarttag-details}. 

The SmartTag GATT profile has four primary services which can be summarized as follows:
\begin{description}
    \item[Authentication Service] 
    The Authentication Service uses  
    three characteristics, \texttt{NONCE}, \texttt{ENONCE} and \texttt{SUPPORTED\_CIPHER}, for authenticating a connected device over BLE. 
    
    \item[DFU Service]
    Service UUID FE59 is a part of the nRF52833 Buttonless Secure DFU service for over-the-air firmware updates. It has a writeable characteristic (\texttt{BUTTONLESS\_DFU}) that can be used to reboot the tag. 
    
    \item[Onboarding Service]
    Service UUID FD59 is used for device onboarding/registration activities. During the registration process, the owner device and the tag exchanges configuration and cryptographic data over various characteristics under this service.
    
    \item[Command Service]
    Service UUID FD5A is primarily used for performing more complicated interactions between the owner device and a tag, such as executing a supported command (e.g., alarm, changing ringtone) on the tag.
\end{description}

A SmartTag registration is initiated by an owner device running the SmartThings app. It involves online interactions with the server, and offline interactions with the tag. The interaction with the server requires a valid user account. In the following discussion, we shall assume that a valid user account has been created and an authenticated session between the owner device and the server has been established.
The registration protocol consists of five stages: key establishment, owner-tag authentication, confirmation of physical possession of the tag, tag ownership check, and registration finalisation.  

\paragraph{Stage 1: key establishment}\label{ssec:shared-secret-establishment}

\begin{figure}[htp]
    \centering
  \includegraphics[width=\columnwidth]{{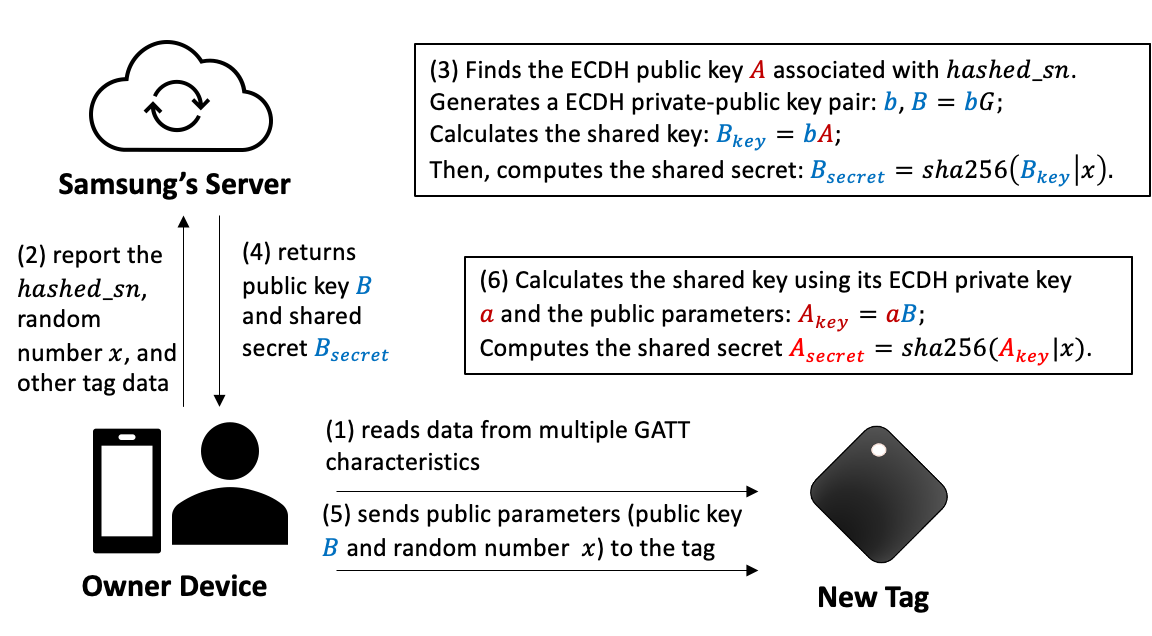}}
  \caption{Shared secret establishment protocol}
  \label{fig:verification-ecdh}
\end{figure}

The first stage of the registration protocol is essentially a key establishment protocol between the tag and the server, mediated by the owner device. Note that since we do not have access to the server code, our analysis of this stage is based on our analysis of the SmartTag firmware and the intercepted traffic between the owner device and the server. Our findings suggest that the tag keeps a pair of private-public ECDH key $(a, A_{pub})$, which is fixed for the lifetime of the tag and that the server keeps at least the public key of each tag. A tag is identified uniquely via its serial number, which is the  identity address of its BLE controller. The public-private key pair of the tag is never sent out from the tag or the server.   
The shared secret establishment protocol is 
summarized in Figure  \ref{fig:verification-ecdh}, which we elaborate below.
\begin{description}[style=unboxed,leftmargin=0.5cm]
    \item[Step 1.] The owner device obtains the necessary registration data from the tag, through the tag's advertisement payload and Onboarding Service. Among this registration data is a hashed serial number (\lstinline$hashed_sn$) unique to the tag, which corresponds to the SHA256 digest of the BLE identity address of the tag. 
    
    \item[Step 2.] The owner device generates a 32-byte random value $x$, and sends $x$ along with the registration data obtained from Step 1 to the server. 

    \item[Step 3 \& 4.]
    After receiving the request, 
    the server looks up the public key of the tag $A_{pub}$ associated with the \lstinline$hashed_sn$. The server then generates an ephemeral private-public key pair $(b, B_{pub})$, and computes the ECDH shared key $B_{key} = bA$. Finally, the shared key is concatenated with the random number $x$ to form the input for the SHA-256 hash function to produce the shared secret: $B_{secret}=SHA256(B_{key}|x)$.
    The server then sends $B_{pub}$ and $B_{secret}$ to the owner device. 
    
    \item[Step 5.] The owner receives $B_{pub}$ and $B_{secret}$ from the server, 
    and forwards $B_{pub}$ and $x$, to the tag. 
    
    \item[Step 6.] The tag receives $B_{pub}$ and $x$ from the owner device and computes $A_{key} = aB$ and $A_{secret} = SHA256(A_{key} | x).$ 
\end{description}

By the property of ECDH, assuming no tampering from an adversary, at the end of Step 6, we should have $A_{secret} = B_{secret}.$ That is, all participants now share the same secret $B_{secret}.$ This shared secret will be used next to compute several AES keys that will be used in subsequent protocols. 

The first 16 bytes of the shared secret $B_{secret}$ are taken as the \texttt{masterSecret}. It is used to derive six 16-byte subkeys for securing communication between the owner device and the tag. 
Note that Samsung OF protocol does not use any default BLE pairing and authentication mechanisms, so this shared secret is unrelated to BLE Long Term Key (LTK) that is normally exchanged as part of BLE pairing protocols~\cite{bluetoothcore}.

The subkeys are derived using a key derivation function, given below, that combines the \texttt{masterSecret} and a parameter that is used to differentiate subkeys: 
\[
\mathit{kdf}(k,x) = SHA256(m(k,x))[0:15]
\]
\[
\mbox{where }  m(k,x) = 
        \begin{tabular}{|c|c|c|} 
        \hline
        Bytes 0-15 & Bytes 16-19 & Bytes 20-\\ 
        \hline
        $k$ & 00000001 & $x$  \\
        \hline
        \end{tabular}.
\]

The following four subkeys are computed by the owner device, the tag and (presumably) the server.
\begin{enumerate}
    \item \textbf{Owner authentication key}: This key is used by the owner to establish an authenticated BLE session with a SmartTag. It is computed by applying : 
    \[
    \mathit{AK}_o = \mathit{kdf}(\text{masterSecret}, \text{"bleAuthentication"})
    \]
    
    \item \textbf{Owner GATT key}: This key is used for encrypting the data exchanged in the GATT interactions between the owner and the tag. It is dependent on a nonce and is valid for a single session of interactions with the tag. 
    \[
    \mathit{GK}_o(\mathit{nonce}) = \mathit{kdf}(\text{masterSecret}, \mathit{nonce})
    \]
    Here $nonce$ is a 16-byte value received from the SmartTag during each BLE authentication process, which will be discussed in \S\ref{sssec:bleauth}.
    
    \item \textbf{Privacy key} \label{itm:pidk}: This key is used for generating unique privacy IDs for a SmartTag (see \S\ref{ssec:lost-and-found}). 
    \[
    \mathit{PIDK} = \mathit{kdf}(\text{masterSecret}, \text{"privacy"}).
    \]
    
    \item \textbf{Advertisement signing key} \label{itm:ask}: This key is used for signing and validating the integrity of the BLE data broadcasted by a SmartTag:
    \[
    \mathit{ASK} = \mathit{kdf}(\text{masterSecret}, \text{"signing"}) 
    \]
\end{enumerate}

Two additional subkeys are derived by a SmartTag and the server, but not the owner device (see \S\ref{ssec:unknown-tag} for details of when and how these keys are used).
\begin{itemize}
    \item \textbf{Non-owner authentication key}. This key is used by a non-owner device to authenticate to the tag.
    \[
    \mathit{AK}_{no} = \mathit{kdf}(\text{masterSecret}, \text{"nonOwner"})
    \]

\item \textbf{Non-owner GATT key}. This key is used in a GATT session between a non-owner and the tag for exchanging commands. It is dependent on a nonce that is exchanged during the GATT interaction.
\[ 
\mathit{GK}_{no}(\mathit{nonce}) = \mathit{kdf}(\mathit{AK}_{no}, \mathit{nonce}). 
\]
\end{itemize}
Notice that unlike the owner GATT key, the non-owner GATT key is not generated directly from the masterSecret; rather it is derived from the non-owner authentication key. 

\paragraph{Stage 2: Owner-tag authentication}
\label{sssec:bleauth}
After computing the \texttt{masterSecret} and the AES keys, the owner device initiates a two-way authentication with the tag to establish an authenticated connected session. This protocol is implemented through BLE interaction only, using the Authentication Service of the GATT profile of the tag. In the protocol description below, $O$ denotes the owner device and $T$ denotes the tag. Throughout the remainder of the paper, we shall use the notation
$E_k(x,y)$ to denote the AES/CBC/PKCS7 encryption of plaintext $y$ with key $k$ and initialization vector $x.$

\[
\begin{array}{llcl}
1. & O \to T & : & n_O\\
2. & T \to O & : & n_T\\
3. & O \to T & : & E_{AK_o}(n_T, \text{"smartthings"})\\
4. & T \to O & : & E_{AK_o}(n_O, \text{"smartthings"})
\end{array}
\]
Here $n_O$ and $n_T$ refer to nonces generated by $O$ and $T$, respectively. At Step 3, the tag checks that the received ciphertext is indeed the encryption of the text \text{"smartthings"}; likewise in Step 4, the owner checks that the received ciphertext is of the expected form. If any of these checks fail, the authentication fails; otherwise the authentication is established, and both the tag and the owner derive an Owner GATT key using $n_T$, i.e., $gk_o = GK_o(n_T).$ This key acts as a session key that is used to secure data transmission in this authenticated session, until the BLE connection is terminated.

\paragraph{Stage 3: confirming physical possession of the tag.}\label{ssec:validation}

After the owner and the tag have successfully established an authenticated BLE connection and derived the session key $gk_o$, the next step is to establish the physical presence of the tag. In a normal registration flow, the SmartThings app will ask the user to press the tag button to ensure physical possession of the tag. 
Pressing the tag button sets the value of the \lstinline$CONFIRM_STATUS$ characteristic on the tag to 0x01 (encrypted using $gk_o$ with IV set to $n_T$) from the default value 0x00. 
\[
T \to O ~:~ E_{gk_o}(n_T, 0x01) \\
\]
The owner's device would only continue the registration flow after validating the value of this characteristic.

\paragraph{Stage 4: tag ownership status check}\label{ssec:ownership-status}
This stage checks the ownership status of the tag to ensure that it is not currently registered to another user. This is done via a simple request to the server, containing the serial number $sn$ of the tag. If the server indicates that the tag has already been registered to another account, the registration process will abort. 
 
\paragraph{Stage 5: finalizing tag registration}\label{ssec:finalization}

This stage creates an online profile of the tag, associated with the owner's account. The protocol can be described abstractly as follows (a more detailed version can be found in Appendix~\ref{sec:smarttag-details}), where $S$ denotes the server. 

\[
\begin{array}{lll}
1. & O \to S: & sn, id, B_{secret}\\
2. & S \to O: & deviceId, metadata\\
3. & O \to T: & E_{gk_o}(n_T, metadata)\\
4. & O \to T: & E_{gk_o}(n_T, curtime)\\
5. & O \to T: & E_{gk_o}(n_T, \text{"Finish"})\\
6. & T \to O: & pid\\
7. & O \to S: & pid, loc 
\end{array}
\]
In Step 1, the owner device sends a record, containing, among others, the tag serial number $sn$, an identifier $id$ and the shared secret $B_{secret}$ established in \S\ref{ssec:shared-secret-establishment}. Recall that $sn$ is used in the ownership status check in the previous stage, to avoid a tag being registered twice. Our observations showed that $sn$ and $id$ were identical and equal to the tag BLE identity address. 

In Step 2, the server returns a unique $deviceId$ that is linked to the tag, and a $metadata$ record that contains the following fields: the {\em privacy Id pool size} ($pidsize$), the {\em privacy Id seed} ($pidseed$) and the {\em privacy Id IV} ($pidIV$). The owner device keeps a record of these parameters and in Step 3, forwards them (encrypted) to the tag. 
These parameters are used later by the tag to generate BLE advertisements. 
The $deviceId$ can be used later to lookup the location of the tag. 

In Step 4, the owner device sends the current time information (in UTC) to the tag so the tag can synchronize its time with the owner. Finally, the owner sends an encrypted message (\texttt{"Finish"}) to indicate the end of the registration process and disconnect from the tag in Step 5.

In Step 6, the tag broadcasts BLE data containing a privacy ID (see \S\ref{ssec:lost-and-found}) that the owner device uses to identify it. After discovering its own tag, in Step 7, the owner device reports the current location $loc$ and the tag $pid$ to the location server. The first location report of a tag made by its owner device creates a device profile on the server, which bonds the tag's ownership status with the owner's Samsung account, preventing others from registering the tag (\S\ref{ssec:ownership-status}). Crucially, the server uses the value $id$ as the identity of the tag for the ownership bonding.

\subsection{Owner-Tag interaction}\label{sec:gatt-commands}
A tag may move in and out of the BLE range of its owner device. The owner device (which also acts as a helper device for tags it does not own) periodically scans for tags in its proximity. When a tag is detected in the proximity of its owner, after having been away, the owner (through the SmartThings app) will automatically initiate the owner-tag authentication protocol over BLE, as described in \S\ref{sssec:bleauth}. 
Upon successful authentication, the app will show the tag as connected and the owner can perform various supported commands on the tag, such as ringing the tag. 
SmartThings also allow an option to configure the tag to perform limited actions on the owner device, such as ringing the owner device when a physical button in the tag is pressed. Appendix~\ref{sec:smarttag-details} contains details of some of these commands. Each command is triggered by exchanging encrypted data $E_{gk_o}(n_T, Data)$, where $gk_o$ is the owner GATT key and $n_T$ is the tag's nonce sent during the owner-tag authentication stage (\S\ref{sssec:bleauth}), and $Data$ is an encoding of the command: 
    \[
      Data = 
        \begin{tabular}{|c|c|c|} 
        \hline
        Bytes 0-3 & 4  & 5-         \\ 
        \hline
        Counter & Opcode & Argument(s)  \\
        \hline
        \end{tabular}
    \]

The counter value corresponds to the total number of commands successfully sent by $O$ during the current authenticated session. The opcode specifies the command type, as a characteristic may support multiple commands, e.g., opcode 0 for alarm off, 1 for alarm on. 
\subsection{Location querying and reporting}
\label{ssec:lost-and-found}
We now discuss the subprotocols for an owner device to query its lost tags from the location server, and for a helper device to report location information of lost tags to the server. 
The location querying protocol consists only of simple POST requests to 
an URL under the \url{api.samsung.com} domain, so we shall omit the details. 
In the following, we focus on subprotocols used for location reporting. 

\begin{figure}[ht]
\centering
\begin{tikzpicture}[font=\ttfamily,every node/.style={scale=0.7},
array/.style={matrix of nodes,nodes={draw, minimum size=7mm},column sep=-\pgflinewidth, row sep=0.5mm, nodes in empty cells,
row 1/.style={nodes={draw=none, fill=none, minimum size=5mm}},
row 1 column 2/.style={nodes={draw}},row 1 column 3/.style={nodes={draw}},row 1 column 5/.style={nodes={draw}},row 1 column 6/.style={nodes={draw}}, minimum size=15mm }]
{\setlength{\tabcolsep}{3em}
\matrix[array] (array) {
0 & 1-3 & 4-11 & 12 & 13-15 & 16-19 \\
15 & 6b fa 00 & c8 40 62 b2 8f 00 e2 60 & c3 & 00 00 00 & ad 01 8b 47 \\};
}
\begin{scope}[on background layer]
\fill[gray!10] (array-1-1.north west) rectangle (array-1-6.south east);
\end{scope}
\draw (array-1-2.north)--++(90:2mm) node [above] (first) {aging counter};
\draw (array-1-3.north)--++(90:2mm) node [above] (first) {privacy ID};
\draw (array-1-5.north)--++(90:3mm) node [above] (first) {reserved};
\draw (array-1-6.north)--++(90:2mm) node [above] (first) {signature};
\node [below=10mm, align=center, anchor=south west] at
(array-2-1.south west) (0) {
\scalebox{1}{
  {\setlength{\tabcolsep}{.2em}
    \begin{tabular}{|c|c|c|c|} 
        \hline
        Bit & 0-3 & 4 & 5-7   \\ 
        \hline
         Info & version & adv type & tag state \\ 
         \hline
        \end{tabular}
        }
    }
    
};
\node [below=20mm, align=center, anchor=south] at
(array-2-4.south) (12) {
\scalebox{1}{
  {\setlength{\tabcolsep}{.2em}
    \begin{tabular}{|c|c|c|c|c|} 
        \hline
        Bit & 0-3 & 4 & 5 & 6-7   \\ 
        \hline
         Info & region ID & E2E flag & UWB flag & battery level \\ 
         \hline
        \end{tabular}
        }
    }
};
\draw (0)--(array-2-1.south);
\draw (12)--(array-2-4.south);
\end{tikzpicture}
\caption{The OF advertisement structure for SmartTags}
\label{fig:advtag} 
\end{figure}


\begin{table}[t]
  \centering
    \begin{tabular}{lll}
      \toprule
      Bits 5-7 & Name  \\
      \midrule
        001 (1) & Premature lost mode  \\
        010 (2) & Lost mode  \\
        011 (3) & Overmature lost mode  \\
        100 (4) & Paired with one device  \\
        101 (5) & Connected to one device \\
        110 (6) & Connected to two devices \\
        \bottomrule
    \end{tabular}
    \caption{Operating states of a SmartTag}
   \label{tbl:smarttag-states}
\end{table}

\paragraph{SmartTag BLE advertisement data.}\label{para:smarttag-ble-adv}
The helper device identifies nearby (lost) SmartTags through their BLE advertisement data only, so we first present details on the advertisement structure and how it is generated. 
Figure \ref{fig:advtag} shows the OF advertisement structure for SmartTags with the following fields highly relevant to the OF protocol: the {\em tag state} (byte 0), the {\em aging counter} (bytes 1-3), the {\em privacy ID} (bytes 4-11), and the {\em signature} (bytes 16-19).  




The tag state is encoded in bits 5-7 of byte 0 and denotes the operating states of a tag.  
There are six different tag states, shown in Table~\ref{tbl:smarttag-states}.  
The state of a registered tag becomes premature lost once it is disconnected from its owner device or rebooted. After operating in premature lost mode for 15 minutes, the state changes to lost, which informs nearby helper devices that it is considered lost. After 24 hours in lost mode, the state becomes overmature lost, where certain BLE operations slow down for power saving. A helper device that finds a tag in the overmature lost mode will initiate an anti-tracking process (see \S\ref{ssec:unknown-tag}). 
    

Bytes 1-3 store the aging counter, a timestamp computed using the tag's time (UTC) and a hardcoded constant: $agingCounter = (tagTime - 1593648000)/900$. 
The aging counter is universal for all SmartTags in-sync with the server's time and updates periodically.


Bytes 4-11 store an 8-byte privacy ID, which is a unique identifier of a SmartTag. Each registered SmartTag has a set of unique privacy IDs called the privacy ID pool. This set is deterministically generated using the privacy key ($PIDK$) 
and the privacy ID configurations ($pidsize$, $pidseed$, $pidIV$) that the server sent to the owner during the finalization stage in the registration process (\S\ref{ssec:finalization}). The privacy IDs are an essential part of the OF protocol as they associate a tag with its owner's Samsung account.
For each $i \in \{1,\dots,pidsize\}$, the privacy ID $pid_i$ is generated as follows: 
\[
    pid_i = E_{PIDK}(pidIV, input_i), \mbox{ where $input_i$ is the byte array} 
\]
\[ 
        \begin{tabular}{|c|c|c|c|c|} 
        \hline
        Byte 0  & Byte 1  & Byte 2-9 & Byte 10 & Byte 11  \\ 
        \hline
        $x_i$ & $y_i$ & $pidseed$ & $x_i$ & $y_i$ \\
        \hline
        \end{tabular}
\]
$x_i=(i\gg 8)\land 256$ and $y_i = i\land 256$.
Here, $\gg$ represents right bitshift and $\land$ denotes bitwise AND.

The privacy pool size for SmartTags is 1000 (for the firmware version we analyzed).

The signature field at the end of the advertisement data serves as a cryptographic checksum for the first 16 bytes. 
Let $blePayload$ denote the first 16 bytes of the BLE advertisement data. Then the four signature bytes are obtained from the first 4 bytes of the $fullSignature$ defined below: 
    \[
      fullSignature = E_{ASK}(pidIV,blePayload)
    \]
where $ASK$ is the advertisement signing key (see subkey \ref{itm:ask}).
Any changes to the first 16 bytes of the advertisement will likely cause the signature bytes to change correspondingly, which allows the integrity of the BLE data to be validated by parties with the privacy ID configuration and the shared secret of the tag, such as the owner and the server.
           

A tag rotates its advertisement data periodically. A tag in any non-overmature lost mode updates its privacy ID, aging counter (incremented by 1), and signature every 15 minutes.
Under the overmature lost mode, a tag updates the aging counter and signature every 15 minutes. However, the frequency for shuffling the privacy ID reduces from once every 15 minutes to once every 24 hours.

\paragraph{Location reporting}\label{para:location-reporting}
A helper device regularly scans for BLE advertisement data from nearby SmartTags. It filters BLE advertisements based on the advertising UUID for SmartTags (FD5A). 
The helper device stores the privacy IDs of lost devices in a local database that can store up to 1000 entries. 
Note that these privacy IDs do not necessarily represent distinct devices, as a tag can generate multiple privacy IDs and the (honest) helper device does not have the privacy key ($PIDK$) needed to link these IDs. 
A privacy ID of a tag is marked as expired if it has not appeared in the BLE scanning for 15 minutes and will be removed from the database. Moreover, the helper device only reports locations of SmartTags in lost or overmature lost mode. 

The helper device will report geolocations of lost SmartTags in the database based on estimated locations received from the WiFi or GPS service. Through reverse engineering and runtime analysis on a helper device, we found that each helper device has a pair of RSA 2048-bit private signing key ($pk_H$) and its corresponding verification key ($pub_H$) stored in the device, secured using Android keystore. Our (limited) experiments show that these keys may not be unique per device, as we identify the same keys used in another Samsung device. The public key $pub_H$ is signed by an intermediate CA owned by Samsung. These keys are used in the location report protocol below to certify the originality of the report (in that it originated from an official Samsung device rather than an unauthorized third party). 

To submit a location report, the helper must first obtain an access token from the server. This token request process is re-used again in another protocol (for interacting with an unknown tag), so we describe it separately here. 
\[
\begin{array}{llcl}
1. & H \to S & : & REQ\\
2. & S \to H & : & n_S\\
3. & H \to S & : & pub_R, n_S, sig(n_S, pk_H), cert(pub_H)\\
4. & S \to H & : & access\_token\\
\end{array}
\]

In Step 1, the helper device generates a pair of private-public key $(pk_R, pub_R)$ and then sends a request for location report to the server $S$. Concretely this step is done via a simple HTTPS GET request to $S.$ The server responds with a 16-byte nonce, encoded as a hexadecimal string. $H$ extracts the signing key $pk_H$ from its keystore, after having performed an attestation protocol, and uses it to sign the nonce $n_S$. The helper then sends the nonce $n_S$, its signature $sig(n_S, pk_H)$, the certified public key $cert(pub_H)$, and $pub_R$ to server. Note that the key pair $(pk_R, pub_R)$ is not used in the location report but will be important later in another protocol. If the server checks the validity of the certificate and verifies the signature and sends a unique access token (which is a JWE token, tied to the nonce $n_S$) (Step 4) if everything checks out. The access token is then used to authenticate the location report. 
\[ 
H \to S ~ : ~ access\_token, report
\]
where $report$ contains a list of advertisement data and the geolocation they were detected. 
Each location report allows a maximum of 5 recently found devices ($time_{found}\geq time_{current} - 1$ (minute)) from the local database to be reported. Each access token has a fixed expiry time (around 32 hours, based on our experiments), after which, the token request protocol above must be repeated to obtain a  new token.

\subsection{Unknown tag detection}\label{ssec:unknown-tag}
The SmartThings app has a feature for detecting nearby overmature lost mode tags for anti-stalking purposes. If such as tag has been detected to be following a helper device, the SmartThings app gives the device owner an option to play sound on the tag to help locating it. Unlike AirTags where any user can issue a command to the tag to play sound~\cite{Heinrich21}, a SmartTag only responds to commands from an authenticated device. 
Since the helper device does not have the authentication key to authenticate to the tag, it would need the vendor server's help. 

We assume that the helper device has performed the protocol to request JWE access token (see the location reporting protocol above) and obtained the token. To play sound on the tag, the helper needs to initiate a GATT connection and reads off a nonce from a certain characteristic. In the following, the notation $AE_{pub}(x)$ denotes an asymmetric encryption of plaintext $x$ with public key $pub$.
\[
\begin{array}{llcl}
1. & H \to T & : & n_H\\
2. & T \to H & : & n_T\\
3. & H \to S & : & access\_token, n_T, pid\\
4. & S \to H & : & X, Y\\ 
5. & H \to T & : & en_T\\
6. & T \to H & : & E_{AK_{no}}(n_H)\\
7. & H \to T & : & E_{gk_{no}}(n_T, command)
\end{array}
\]
In Step 4, the messages are computed as follows:
\[
X = AE_{pub_R}(E_{AK_{no}}(n_T)) \qquad Y=AE_{pub_R}(GK_{no}(n_T))
\]
where $pub_R$ is the public key $H$ generated in the access token request protocol. $H$ decrypts these to obtain $en_T = E_{AK_{no}}(n_T)$ and $gk_{no} = GK_{no}(n_T)$ used in the remainder of the protocol. 
If we omit step 3 and 4, this protocol is similar to the owner-tag authentication protocol, except that the authentication key used is $AK_{no}$, and the GATT key used to send the command is $GK_{no}(n_T)$. 
Note that the server never discloses the $AK_{no}$ itself. This request needs to be accompanied by a current access token, and an identifying information from the tag (i.e., its privacy ID $pid$, and some other information we omit here).

\section{Security and privacy analysis}
\label{sec:analysis}
The four research questions guide the construction of our adversary model (Table \ref{tbl:adversary-model}) which we use for our analysis. This model classifies potential threats to the OF system into four categories based on our research objectives, extending  the FindMy adversary model proposed in \cite{Heinrich21}. Firstly, our model subdivides proximity-based threats (A1) into passive (A1.1) and active (A1.2) categories, based on the level of interaction with the SmartTag's BLE interfaces.
In line with the FindMy model, we consider network-based (A2) and service operator (A3) threats. Finally, we introduce a new Tag Owner category (A4), acknowledging potential security implications posed by SmartTag owners themselves. The final column indicates the related research questions, 
highlighting the implications of each threat category.
\begin{table*}[htp]
  \centering
  \caption{Adversary Model for Samsung SmartTags}
  \label{tbl:adversary-model}
{\small %
    \begin{tabular}{|p{.24\textwidth}|p{.31\textwidth}|p{.3\textwidth}|p{.05\textwidth}|}
      \hline
      Model & Assumptions & Capabilities & Impact \\
      \hline
      Passive Proximity-based (A1.1) &
      \multirow{2}{=}{Within BLE communication distance with a tag; Controls a Bluetooth device} &
      Record and replay BLE advertisements &
      \multirow{2}{=}{RQ1 (\S\ref{para:attack-scenario-A1})} \\
      \cline{1-1}\cline{3-3}
      Active Proximity-based (A1.2)
       &
       &
      Interact with tag’s GATT server &
      \\
      \hline
      Network-based (A2) &
      MITM position between Samsung server and a tag. &
      Intercept, redirect, or modify network traffic &
      RQ4 (\S\ref{para:attack-scenario-A2}) \\
      \hline
      Service Operator (A3) &
      Access to backend systems. &
      Access to all location reports and secret keys for each registered SmartTag. &
      RQ3 (\S\ref{para:attack-scenario-A3}) \\
      \hline
      Tag Owner (A4) &
      Owns a SmartTag; Controls a Bluetooth device; Direct contact with a victim &
      Hide the tag/customized tracking device in victim's belongings &
      RQ2 (\S\ref{para:attack-scenario-A4}) \\
      \hline
    \end{tabular}%
}%
\end{table*}

We now analyze the security and privacy issues affecting Samsung's OF system in alignment with our research questions, each addressed in the following sections: \S\ref{sec:privpool} (RQ1), \S\ref{sec:forging-loc-report} (RQ4), \S\ref{sec:end-to-end-priv} (RQ3), and \S\ref{sec:unwanted-tracking} (RQ2).
Then, we provide an update on various bugs discussed in Section~\ref{sec:bug-fixes}. 

\subsection{Proximity-Based Attack (A1-RQ1)}
\paragraph{Attack Scenario}\label{para:attack-scenario-A1}
By passively eavesdropping on BLE advertisements (A1.1) or actively engaging with the SmartTag’s GATT server (A1.2), the attacker could identify and track the presence of the neighbour's FMM device or SmartTag, thereby gaining insights into their daily schedule.

\paragraph{Linkability Flaws}\label{sec:privpool}\label{sec:fingerprint}
The privacy ID pool size for OF devices is specified through a parameter sent by the server during the registration phase. For FMM devices, this is 51, while for SmartTags, it is 1000. As the pool size is not hardcoded on the client side, this could change in future updates to the server. As it stands currently, even with 1000 pool size, there is a high probability (due to the birthday theorem) that the same privacy ID could be reused in approximately $\sqrt{n}$ rotations, where $n$ is the pool size.  
From our experiments, a few days of passive observations of the BLE data would cover all 51 privacy IDs of an FMM device. 
Recent versions of the FMM app have added a layer of obfuscation to the privacy IDs, keeping the same privacy ID pool size. However, our preliminary finding shows this can be easily de-obfuscated (see Appendix~\ref{sec:privID-obf}). 
This obfuscation is not implemented in SmartTags.  

\emph{GATT server leaking sensitive data.}
    A registered SmartTag advertises on RPAs that frequently change. However, we have found that sensitive information leaked by characteristics in SmartTags' GATT profile (see Figure \ref{fig:gatt} for details) provides two ways for an attacker to de-anonymize the tag's identity: The \lstinline$IDENTIFIER$ characteristic contains its identity MAC address, and the \lstinline$HASHED_SERIAL_NUMBER$ characteristic contains the SHA256 hash of the identity address. Both values are static and unique for each tag and readable, which can be used by any nearby adversaries to identify the tag.
    
    The \lstinline$SUPPORTED_CIPHER$ characteristic is readable and writable. It contains a default value \texttt{"AES128/CBC/PKCS7"} that specifies the cipher being used for BLE authentication. However, we discovered that writing custom values to this characteristic would overwrite the default value until the tag is restarted. Hence, an adversary can identify a registered tag by writing a custom identifier to this characteristic.

\emph{DFU device reboot.}\label{ssec:dfu}
The Galaxy SmartTag has a DFU (Device Firmware Update) Service for secure over-the-air Firmware updates. We discovered that a SmartTag can be rebooted to the DFU mode by enabling indication on the Buttonless DFU characteristic and writing byte 0x01 to it~\cite{nrfsdk}. This is actually not part of Samsung implementation; rather it is a default service available to nRF52 chipset for firmware updates so this vulnerability is not specific to SmartTags. 

In the DFU mode, the tag advertises on a random static MAC address and waits for new firmware packages. If no data is received over a short period (approximately two minutes), the tag will reboot and resume its lost mode BLE operations. Additionally, the DFU mode reveals the identity MAC address of the tag - the MAC address used in the DFU mode ($addr_{DFU}$) equals to the identity MAC address ($addr_{Id}$) plus one~\cite{nrfsdk}, e.g., if $addr_{DFU}$ is observed to be 11:22:33:44:55:66, it can be inferred that $addr_{Id}$ is 11:22:33:44:55:65. 
However, after coming out of the DFU mode, the aging counter in the OF data is reset to 0, making OF tracking unavailable for the tag since the aging counter will be considered as too old and its location will not be accepted by the server.

\emph{Unintended pairing with a SmartTag.}
    The update to firmware 1.02.06 introduces a new vulnerability. 
    A SmartTag with this new firmware appears to accept pairing request, using the Just Works association mode, allowing the attacker to obtain the IRK and the identity address of the tag. The IRK can then be used to resolve the RPAs the tag use in BLE advertising. The IRK appears to be persistent across reboot and account switching. So removing the tag from a Samsung account and pairing with another account does not reset the IRK. The possession of the IRK allows a more stealthy tracking of the tag, as the attacker does not need to connect to tag; they simply observe the RPAs used to advertise the payload and de-anonymize them using the IRK.

\emph{Reflection Attack.}
        The BLE authentication procedure is a two-way challenge-response protocol, which assumes that the tag $T$ and the owner device $O$ would exchange randomly generated challenges ($n_T$ and $n_O$), followed by the responses (see \S\ref{sssec:bleauth}). This protocol is vulnerable to a reflection attack, where an adversary $A$ can manipulate the Owner Device into providing the answer to its own challenge:
        \[
        \begin{array}{llcl}
        1. & O \to A & : & n_O\\
        2. & A \to O & : & n_O\\
        3. & O \to A & : & x = E_{AK_o}(n_O, \text{"smartthings"})\\
        4. & A \to O & : & x
        \end{array}
        \] 
    In step 2, instead of randomly generating its own challenge, $A$ replays $n_O$ from step 1, and hence the response in step 3 is also the response for $A$'s challenge. This attack would allow an adversary to impersonate a registered SmartTag that can authenticate to tag's Owner Device. It can be prevented by adding a validation mechanism on the $n_T$ received by the Owner Device. For instance, the authentication flow would only continue if $n_T \neq n_O$.

\emph{Unintended Silent Pairing with an Owner Device.}
It is possible to impersonate a SmartTag and silently pair with its owner device by exploiting the following pairing behavior in the BLE specification: if a central device encounters an "Insufficient Authentication" error when interacting with a characteristic in a peripheral, it will initiate the pairing procedure with the peripheral. This behavior has been exploited in previous work~\cite{Xu19NDSS,Zhang20Usenix,Tiu20} to initiate unintended pairing. As noted in \cite{Zhang20Usenix}, the attacker can influence the association method for the pairing, e.g., force the pairing to use a less secure method, such as Just Works.

In this attack, the attacker acts as the peripheral, while the central device is the owner device. The attacker creates a GATT profile of a SmartTag and replay the latest BLE advertisement from the tag to trick the owner device into initiating the BLE Authentication procedure (\S\ref{sssec:bleauth}). BLE Authentication starts with the owner device writing to the \texttt{NONCE} characteristic of the (impersonated) SmartTag's GATT server. By setting the write permission for the \texttt{NONCE} to \texttt{encrypted-write}, the "Insufficient Authentication" error will be triggered upon write requests. 
Prior to the November 2020 patch~\cite{anroidsec}, pairing is performed silently on most Android versions if Just Works is used~\cite{Tiu20}. Older models of Samsung devices that are not eligible to receive the update, such as Samsung Galaxy 7, remain vulnerable.
This attack allows the attacker to obtain the IRK and the identity address of the owner device that can be used for long-term tracking.

\subsection{Network-based Attack (A2-RQ4)}
\label{ssec:network-based}
\begin{figure}[t]
  \centering
  \includegraphics[width=0.8\columnwidth]{{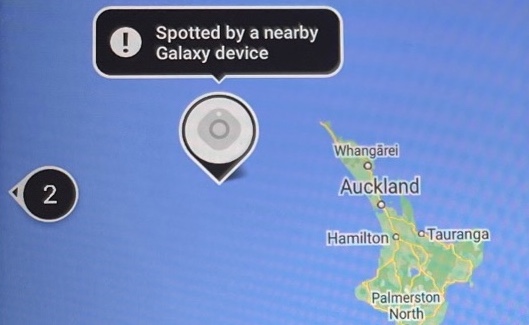}}
  \caption{Fake location being updated on the owner device}
  \label{fig:fake-location}
\end{figure}

\paragraph{Attack Scenario}\label{para:attack-scenario-A2}
In the context of network-based threats, an attacker stealing a tag (attached to a stolen item) may be able to hide the location of the stolen item by forging location reports using the tag's lost mode advertisement, leading the owner to a false trail.

\paragraph{Forging location reports}
\label{sec:forging-loc-report}
We found several ways in which the integrity of the location report can be compromised, allowing an attacker to report fake locations of a lost tag.

\emph{Relay attack.} Recall that the helper device simply forwards the BLE advertisement data of a lost tag or mobile device to the location server. There are no mechanisms for the helper to determine whether the data was indeed broadcasted by a nearby legitimate OF device. This allows a very simple relay attack: two attackers $A$ and $B$ in two different locations can collude by forwarding the BLE advertisement data of a device observed in $A$'s location to $B$ to be replayed at $B$'s location, and vice versa. If there are helper devices in both locations, they will submit conflicting reports. Indeed, our experiments show the location server does not check for the consistency of the location reports, e.g., one device could appear to be detected in two different continents within seconds. This attack is especially effective against FMM devices as their advertisement data has no expiry time and can be replayed indefinitely. In contrast, with SmartTags, the aging counter in the advertisement prevents an indefinite replay. From our observation, we've noticed that advertisement data that is more than 7 days old will be rejected by the server.

Notably, this appears to be a universal issue affecting all crowd-sourced BLE tracking systems, including Apple's FindMy. Potential mitigation will be discussed in \S\ref{sec:discussions}, such as using a distance-bounding protocol.
However, we emphasize that this is a difficult problem that extends beyond the scope of this paper.

The next three attacks exploit the location report protocol itself. They correspond to three different ways in which the attacker can obtain an access token to submit location reports (see \S\ref{ssec:lost-and-found}). Recall that the location report will only be accepted by the location server if the reporter has a valid access token. These attacks are more powerful as they do not require any helper devices to be present at all at the target location. For example, we were able to forge a report of an OF device detected in the middle of an ocean, as shown in Figure \ref{fig:fake-location}. 

\emph{Re-using location reporting access token.}
As it turned out, once the helper device receives an access token from the vendor's server for location reporting, it records the token in its system log. Consequently, an attacker who owns a helper device can extract this token from the log and send fake location reports using it. This process can be easily automated as the report is done through a simple HTTPS POST request to the location server. 

\emph{Access token renewal through signature replay.} 
We discovered that in the access token request protocol, the server does not check whether the nonce $n_S$ it sent in Step 2 is the same as the $n_S$ it receives in Step 3. This allows an attacker who is in possession of a signed nonce from a previous session to replay it to get another access token. More specifically, suppose the attacker has $n_S'$ and $sig(n_S', pk_H).$ Then the following is a valid protocol run for token renewal: 
\[
\begin{array}{llcl}
1. & H \to S & : & REQ\\
2. & S \to H & : & n_S\\
3. & H \to S & : & pub_R, n_S', sig(n_S', pk_H), cert(pub_H)\\
4. & S \to H & : & access\_token\\
\end{array}
\]
even when $n_S \not = n_S'.$ Note that as all communication between the helper device and the location server are secured via HTTPS, for this attack to be possible, the attacker would need to be able to decrypt the TLS encrypted traffic. In our experiments, this was done by installing a root CA in a rooted helper device, and perform MITM attack between the device and the server. 
We observed that in recent Samsung phones running Android 10 or later, the location report process involves an attestation protocol before the phone could extract the signing key to sign the nonce. This attestation step would fail if it detects the phone is rooted. We only managed to execute this attack in an older Samsung phone running Android 8.0. 

\emph{Extracting the signing key.} 
We discovered a flaw that allowed us to extract the signing key itself. Proof for the extracted signing key can be found in Appendix \ref{sssec:apx-smarttag-location-reporting}.
This then allowed us to run the entire location reporting protocol outside the phone as we now possess all the information to pass the authentication stage to get the access token. 

\subsection{Service Operator (A3-RQ3)}
\label{sec:end-to-end-priv}
\paragraph{Attack Scenario}\label{para:attack-scenario-A3}
Without a strong end-to-end privacy, the service operator may infringe user privacy by inferring social connections through location history analysis.

\paragraph{End-to-end privacy}
The OF protocol assumes the vendor as the trusted party, since the vendor has the key material needed 
to compute the privacy IDs for any registered SmartTag.
This means that the vendor can de-anonymize the location reports of a lost device. From a privacy standpoint, this is worse than Apple's FindMy network, in which the cryptographic key needed to generate the privacy IDs is not disclosed to the vendor. 

A more interesting question is whether the design of Samsung OF protocol protects the privacy of helper devices. From the location reporting protocol in \S\ref{ssec:lost-and-found}, we see that the access token can be used to link multiple location reports. Since each access token is assigned to a helper device, its location reports can be linked to plot its trajectory over the validity period of the access token. In principle, the access token is not tied to a particular Samsung account. However, under normal operations, such a token request would be accompanied by other requests to Samsung servers that they can potentially be correlated. 

\subsection{Tag Owner (A4-RQ2)}
\paragraph{Attack Scenario}\label{para:attack-scenario-A4}
Without adequate anti-tracking mechanisms, malicious SmartTag owners could misuse the OF network for stalking purposes, e.g., covertly track a colleague by hiding the tag in their belongings, or create a hard-to-detect customized tracker leveraging Samsung's OF protocol. 

\paragraph{OF device emulation.} 
\label{sec:impersonation}
Emulating an OF device helps both in understanding the various sub-protocols involved in the interactions between the devices and vendor's server and in evaluating the feasibility of creating custom trackers to evade Samsung's anti-tracking mechanisms. 
Emulating FMM mobile devices is straightforward as the (only) secret for generating privacy IDs can be extracted easily from the device log. 
For SmartTags, this is not possible. To impersonate a SmartTag that can be registered through normal flow, the impersonation would need to successfully pass the finalization stage (see \S\ref{ssec:finalization}), to establish a shared secret with the server. Since the private key for deriving the secret is embedded in the hardware of a legitimate tag and is not explicitly exchanged during registration, it cannot be obtained easily without performing a hardware-level attack on the tag. Unlike the case with FMM device, this shared secret is not recorded in the device log of the owner device. 
However, we managed to obtain the shared secret by setting up a MITM attack between the owner device 
and the server, and monitor their exchanges. Since this shared secret is sent by the server to the owner device, we sidestep the need for extracting the private-public key of the tag.

\paragraph{Anti tracking feature}
\label{sec:unwanted-tracking}
The FMM version 7.2.25.14  introduces an anti-tracking module for background detection of tracking tags. The OF data advertised by a lost mode SmartTag contains two temporary identifiers: the Non-Resolvable RPA it uses and the privacy ID in the BLE payload. For an lost mode tag, the two identifiers change every 15 minutes. For an overmature lost mode tag, the MAC address still randomizes once every 15 minutes, yet the privacy ID only updates once every 24 hours. 

We noticed that the anti-tracking feature could only detect SmartTags in overmature lost mode through initial observations. Reverse-engineering of the FMM app has shown that the tracking detection algorithm uses the privacy ID contained in the BLE advertisement data of a tag as its identifier, then uses two thresholds to determine whether the tag is a tracker: (1) the duration since the tag was first discovered and saved to the local database; (2) the distance traveled while the tag is nearby (according to the geolocations saved to the database).
This explains why the algorithm cannot detect tracking tags in lost mode: Since the anti-tracking algorithm uses the privacy ID data as the only identifier of a tag, it cannot correlate a tag before and after its privacy ID changes. Thus, the detected tracking duration is at most 15 minutes for an lost mode tag. While, for an overmature lost mode tag, the privacy ID value only changes once a day, which typically allows both thresholds to eventually satisfy. Another consequence of this is that privacy IDs generated by offline FMM mobile devices are also ignored by the anti-tracking mechanism.

The above allows an attacker to circumvent anti-tracking through a custom BLE tag that either impersonates an offline FMM device, or a SmartTag with fast rotating privacy IDs. 

\subsection{Updates on the vendor bug fixes}
\label{sec:bug-fixes}

\begin{enumerate}
  \item Small Privacy ID Pool (\S\ref{sec:privpool}): The pool size remains the same, but an obfuscation process has been introduced to make the privacy IDs appear more unique. However, attackers can de-obfuscate these IDs (Appendix \ref{sec:privID-obf}).
  \item Linkability Flaws (\S\ref{sec:fingerprint}): In the latest firmware version (1.04.01), the \lstinline$IDENTIFIER$ characteristic only returns the tag's serial number to authenticated devices. 
  the \lstinline$HASHED_SN$ is no longer writable, and the silent pairing (with tag) vulnerability has been fixed. However, the DFU service remains vulnerable. 

  \item Location Report Protocol: Samsung has implemented a consistency check to secure the nonce used for access token renewal, preventing the signature replay attack from \S\ref{sec:forging-loc-report}. Yet, the other two location report vulnerabilities remain. 

  
  
\end{enumerate}

\section{Discussions}
\label{sec:discussions}
We now discuss a broader implication of our findings, in terms of what we think are issues that go beyond the specific implementation we presented in the preceding sections. Again, our discussion here is organised around the four research questions we pose in the introduction, which we think highlight important features of an offline location tracking system. We shall contrast Samsung's OF protocol design and implementation against Apple's Find My (as analyzed in \cite{Heinrich21} -- so this may not reflect the latest version). 

Both Apple and Samsung seem to strike a balance in providing privacy to the owner of a tracker (RQ1) while at the same time providing means to detect unwanted tracking (RQ2). The latter seems to be a focus of much public attention, as indicated by articles in major news outlets such as~\cite{airtagstalking}, and has prompted an initiative~\cite{ietf-unwanted-trackers} by Apple and Google to jointly develop a standard to ensure that future BLE trackers will have features built-in that allow cross-platform detection of unwanted tracking. 
On the issue of unlinkability of BLE data, our conclusion, given our findings and many in BLE fingerprinting work~\cite{Celosia19,Celosia19b,Celosia20,Martin19}, it seems quite impossible to design a system that would fully enforce unlinkability. 

On the issues prompted by RQ2 (unwanted tracking), both Apple and Samsung seem to adopt a similar approach: a lost tracker will transition to a ``lost mode'', under which its privacy ID rotates infrequently, allowing an easy detection by a victim's phone. The essential difference in their implementations is in terms of how long it takes for a lost tracker to transition into the lost mode. In Samsung's case, it takes 24 hours, which provides ample time for the attacker to achieve its target. However, both Apple~\cite{Mayberry21} and Samsung designs share a similar flaw: their anti-stalking algorithms seem to ignore BLE devices which are not dedicated trackers. 

Samsung's protocol design follows a fundamentally different approach when compared to Apple's with respect to end-to-end privacy (RQ3). Apple's design follows what we call a {\em decentralised} approach, where the cryptographic keys controlling the generation of privacy IDs are controlled by the end users. This guarantees, in principle at least, that Apple cannot de-anonymize a privacy ID on its own without additional information. This is in contrast to Samsung's {\em centralised} approach, where the cryptographic keys for generating privacy IDs are known and controlled by Samsung server. This fact may come as a surprise to the reader, and as far as we know, it is not something that is well-understood by the public. This centralised approach does, however, have an advantage over the decentralised approach when it comes to identifying a stalking tag. In the case of Apple's AirTags, this identification is done through reading its serial number, which can be linked to the owner's account. But this assumes the victim can locate the tag and access it physically, and that the tag is a genuine AirTag, so not a custom-modified tag (e.g., using the Open Haystack framework~\cite{seemooopenhaystack}). In the case of Samsung's OF system, the server possesses the information to identify a stalking tag through its privacy IDs, without needing physical access to the tag. From a law enforcement perspective, this allows an easier attribution of the attack.  

Lastly, concerning the location report integrity (RQ4), there is an inherent difficulty in preventing the relay attack that is not specific to offline location tracking systems. This issue, for example, also manifests in BLE-based contact-tracing systems. A potential solution could be to adopt some sort of distance-bounding protocols~\cite{BrandsC93} to ensure physical proximity, but whether such an approach is practical and whether it will not introduce further vulnerabilities into the system, is something that is beyond the scope of the current paper. 

\section{Conclusion}
\label{sec:conclusion}

In this work, the Offline Finding (OF) and device management protocols for Find My Mobile (FMM) devices and SmartTags have been thoroughly analyzed, and a security and privacy analysis was performed. Our analysis of the protocols' design and implementation has identified several flaws, allowing each of the research questions to be answered definitively. 

We have also discovered vulnerabilities outside the scope defined by the proposed research questions, including multiple other flaws related to the GATT server implementation for SmartTags, and the flaw in the registration protocol that allows an attacker to register a SmartTag of someone else without knowing its ECDH private key (Appendix \ref{sssec:registration-attacks}). 

Most of the flaws we identified have been fixed in the latest firmware, so 
some of our findings and analysis results may not apply to devices/tags with higher version numbers. However, at the time of writing, our tests show that devices or tags with older firmware/software versions can still participate in the OF network. 
Existing users of SmartTags and other FMM devices who have the option to upgrade the firmware/apps on their devices are encouraged to do so to mitigate some of the issues we discuss here. 

Among the issues we discussed, of great concern is the possibility of unwanted tracking using SmartTags and similar trackers, such as AirTags and Tile trackers, or custom trackers leveraging on these offline finding networks. The current fragmented approach to anti-stalking features leaves a significant number of people vulnerable to unwanted tracking without an effective mean for detecting it. Fortunately, a standardisation effort is on-going, by Apple and Google, to allow a cross-platform detection of unwanted tracking~\cite{ietf-unwanted-trackers}. We hope our analysis would help inform the design choices in such a standardisation.  
For future work, we plan to investigate ways to detect unwanted tracking that are effective against a variety of OF networks, leveraging on existing efforts such as AirGuard~\cite{airguard}.

\bibliographystyle{plain}
\bibliography{reference}

\newpage

\appendix

\section{Table of Acronyms}
\label{sec:acronyms}

\begin{table}[htp]
\begin{tabular}{|l|p{5cm}|}
\hline
\multicolumn{2}{|c|}{\textbf{Nomenclatures}} \\ \hline
\textbf{Acronym} & \textbf{Full Name} \\ \hline
\multicolumn{2}{|c|}{\textit{Common Acronyms}} \\ \hline
AES & Advanced Encryption Standard \\ \hline
API & Application Programming Interface \\ \hline
APK & Android application Package \\ \hline
BLE & Bluetooth Low Energy \\ \hline
CA & Certificate Authority \\ \hline
DFU & Device Firmware Update \\ \hline
ECDH & Elliptic-curve Diffie–Hellman \\ \hline
GAP & Generic Access Profile \\ \hline
GATT & Generic Attribute Profile \\ \hline
GUI & Graphical User Interface \\ \hline
HTTPS & Hypertext Transfer Protocol Secure \\ \hline
IRK & Identity Resolving Key \\ \hline
LTK & Long Term Key \\ \hline
MAC & Media Access Control \\ \hline
MITM & Man-in-the-Middle \\ \hline
RPA & Random private address \\ \hline
TLS & Transport Layer Security \\ \hline
URL & Uniform Resource Locator \\ \hline
UUID & Universally Unique Identifier \\ \hline
\multicolumn{2}{|c|}{\textit{Samsung's Offline Finding Protocol Acronyms}} \\ \hline
FMM & Find My Mobile \\ \hline
Helper Device & A device that discovers and reports lost FMM devices/SmartTags\\ \hline
Lost Device/Tracker & An FMM device/a SmartTag operating in lost mode or overmature lost mode \\ \hline
Owner Device & A device signed in with a Samsung account that owns FMM device(s)/SmartTag(s)\\ \hline
OF & Offline Finding \\ \hline
Privacy ID & A unique identifier of an FMM device/a SmartTag \\ \hline
\end{tabular}
\caption{List of Acronyms}
\label{tab:nomenclatures}
\end{table}
\FloatBarrier

\section{Offline Finding protocol for smartphones}
\label{sec:fmm}\label{sec:fmm-details}

In this section, we discuss our findings in the reverse engineering of offline finding features of the FMM app. The results of this section apply to all versions of FMM app (with the offline finding features) prior to version 7.2.24.12 (July 2022). We have not studied comprehensively the patched FMM app, but some preliminary findings related to how the advertisement payload is generated is given in Appendix~\ref{sec:privID-obf}. 

The OF protocol has multiple modes of operations that depend on the functions supported by the devices involved. 
We outline the main OF protocol that applies to mobile devices, 
which consists of four main operations: Device/Account Registration, lost (offline) device Operation, helper (online) device Operation, and Device/Account Deregistration. 
To simplify presentation, we omit the detailed concrete details of various messages exchanged and would refer to some important data symbolically. We also conflate the different servers used in the OF protocol to a single entity, which we simply refer to as "the server". 

The protocol can be summarised as follows. Initially, devices must complete the registration process with the server to obtain various parameters that will be used in the offline operation. When a registered device goes offline, it starts advertising a unique payload that identifies itself. This payload is picked up by nearby online (registered) devices which parse the payload extracting the device’s identifier. The online device then accesses available location services to find out its own location. It then sends the lost device’s identifier and the location through to the server. The owner of the lost device can then access the server to find out its location. Further details of each operation are outlined below.

\subsection{Device registration}
The Offline Finding (OF) feature can be enabled on Samsung smartphones in the device settings when the device is signed in with a Samsung account. The device will then make an HTTPs request to the \lstinline$/v1/kms/cf/device/registerDevice$ URL of the \lstinline$samsungdive$ server to register itself to the OF network. The request contains the following information:

\begin{itemize}
    \item Secret key: 16-byte random generated by Java Random.
    \item Device ID: Base64 encoding of an MD5 hash of the device’s IMEI, which is constant and unique for a device.
    \item User ID: A value associated with the Samsung account logged in.
    \item Other information such as device type, region code, client version, and device model. 
\end{itemize}

The server then responds with several pieces of information; the most important one (from a security/privacy perspective) is the \lstinline$PrivateIDConfig$, which consists of a 16-byte secret key, an IV (\lstinline$PrivacyIV$), and a size parameter (\lstinline$PrivacyPoolSize$) that will be used to generate BLE advertisements for the OF operations. The IV value is fixed to the same 16-byte value (i.e., \texttt{"+IABCfvBZHJYFUek8vp3Gg=="} in base64 encoding)  across accounts and devices. The privacy ID pool size determines the amount of possible advertising values the device will generate. This has also been observed to be standard for all devices, taking the value 51. This means that there are only 51 possible advertisement payloads from a lost device.

\paragraph{Registration process details}
Listing \ref{lst:fmm-v1-registration} contains the system log produced when enabling the Offline Finding Service on a phone. The Highlighted parts show the information for the device registration request and response. The request contains the account and device information, such as \lstinline$userId$,  \lstinline$deviceId$, and \lstinline$email$. The server's response contains the Offline Finding policy, the \lstinline$privateIdConfig$, and other configuration data.

\begin{lstlisting}[frame=none,style=base,basicstyle=\scriptsize,numbers=none,linewidth=\columnwidth,caption={FMM device registration log},label={lst:fmm-v1-registration}]
10-07 17:35:33.421  8983  8983 I DBG_FMESEC :[OfflineFindingService]: OfflineFindingService onStartCommand
10-07 17:35:33.421  8983  8983 I DBG_FMMSEC :[DeviceUtil          ]: isOfflineFindingEnabled : false
10-07 17:35:33.422  8983  8983 I DBG_FMESEC :[CFTaskExecutor      ]: executeTask : CFDeviceRegisterTask
10-07 17:35:33.429  8983 21707 I DBG_FMMSEC :[AccountManager      ]: getAccountInfo from preference : `AccountVO{userId='***', deviceId='IMEI:', cntCode='AUS', email='***', serverUrl='www.ospserver.net', mcc='505', remoteControl=true}`
10-07 17:35:33.430  8983 21707 I DBG_FMESEC :[CFHTTPRequestor     ]: request url : https://eu-kms.samsungdive.com/v1/kms/cf/device/registerDevice
10-07 17:35:56.916  8983 21707 I DBG_FMESEC :[CFHTTPRequestor     ]: Reqeust result : 
    `{
       "policy":{
          "version":"1",
          "type":"mobile",
          "advertiseInterval":10000,
          ...
       },
       "TargetURL":"***",
       "privateIdConfig":{
          "secretKey":"***",
          "iv":"+IABCfvBZHJYFUek8vp3Gg==",
          "privacyIdPoolSize":51
       },
       "responseCode":200
    }`
\end{lstlisting}

\subsection{Lost device operation}
When a OF registered device no longer has an active network connection, it enters ‘Lost Mode’ and triggers the Offline Finding service to start. The lost device then creates a GATT server profile and starts advertising on the main OF service UUID (FD69). The advertising is the fundamental operation for lost devices as part of the main OF protocol. The GATT server is not directly used by the main OF protocol, however it can be interacted with via BLE and is used as part of secondary OF protocols. 

\subsubsection{GATT server profile}
\begin{table}[htp]
\centering
\caption{FMM GATT Services}
\label{tbl:fmm-services}
    \resizebox{.9\columnwidth}{!}{%
    \begin{tabular}{lll}
      \toprule
       Name & UUID \\
      \midrule
        ENCRYPTION &  EEDD5E73-6AA8-4673-8219-398A489DA87C\\
        FME & 4EBE81F6-B952-465E-9ECE-5CA39D4E8955\\
      \bottomrule
    \end{tabular}}
\end{table}
The GATT profile of a lost device contains two services: the encryption service and the FME service, as shown in Table~\ref{tbl:fmm-services}.
\begin{itemize}
\item {\bf Encryption Service} (\texttt{ENCRYPTION}). 
This service is used to implement a challenge-response protocol for authenticating a device that wants to connect to the lost device. It contains three characteristics (see Table~\ref{tbl:char-auth}):
    \begin{table}[htp]
    \centering
    \caption{Characteristics under the Encryption Service}
    \label{tbl:char-auth}
        \resizebox{.9\columnwidth}{!}{%
        \begin{tabular}{lll}
          \toprule
           Name & UUID \\
          \midrule
            NONCE & A12BE31C-5B38-4773-9B9D-3D5735233A7C\\
            ENONCE & 4EBE81F6-B952-465E-9ECE-5CA39D4E8955\\
            SUPPORTED_CIPHER & 50F98BFD-158C-4EFA-ADD4-0A70C2F5DF5D\\
          \bottomrule
        \end{tabular}}
    \end{table}
The \texttt{SUPPORTED\_CIPHER} characteristic is readable and contains information about the cipher to be used, which is AES/CBC/PKCS7. The \texttt{NONCE} characteristic is readable and returns an IV to be used during encryption. This IV is a random nonce that is generated using Java \texttt{SecureRandom} each time a client connects to the server. The \texttt{ENONCE} characteristic is writeable and expects to receive an encrypted version of the string ``smartthings”. This string must be encrypted using the given IV and with the device’s secret key (from the \texttt{PrivateIDConfig}). Writing the correct ciphertext to the encrypted nonce characteristic completes the handshake between client and server.

\item {\bf FME Service} (\texttt{FME}). 
    After completing the handshake, the client is now authenticated and can interact with the characteristics in the FME service. The device’s alarm can be set to ring by writing the byte 01 (encrypted using the same cipher) to the \texttt{ALARM} characteristic (see Table~\ref{tbl:char-fme}).

    \begin{table}[htp]
    \centering
    \caption{Characteristics under the FME Service}
    \label{tbl:char-fme}
        \resizebox{.9\columnwidth}{!}{%
        \begin{tabular}{lll}
          \toprule
          Name & UUID \\
          \midrule
            ALARM & 4a1351bb-d617-4612-a8e3-8dee6ca13e7b\\
            CCCD & 00002902-0000-1000-8000-00805f9034f0\\
            MCF & 0487d871-d55e-44aa-8318-4faa721278e5\\
          \bottomrule
        \end{tabular}}
    \end{table}
\end{itemize}


\subsubsection{BLE operations}

\paragraph{Privacy ID generation}
The lost mode advertisements are the fundamental part of the OF protocol. The lost device generates an advertisement containing a unique identifying payload which is picked up by a helper and reported to Samsung. A key component of the advertisement payload is the {\em privacy ID} that identifies the device uniquely. A device can generate a number of privacy IDs, depending on the privacy ID pool parameter in \texttt{PrivateIDConfig}. 

Let $k$, $iv$ and $p$ denote, respectively, the secret key, IV and the privacy ID pool from the device’s \texttt{PrivateIDConfig}. To generate a privacy ID, first we compute a 20-byte array: 
\[
x_i = \mbox{
    \resizebox{.8\columnwidth}{!}{%
    \begin{tabular}{|c|c|c|c|c|} 
    \hline
    Byte 0 & Byte 1 & Bytes 2-17 & Byte 18 & Byte 19         \\ 
    \hline
    00 & $i$ & $k$ & 00 & $i$ \\
    \hline
    \end{tabular}
    }
}
\]
where $i$ is a 1-byte random nonce in $\{1, \dots, p\}$, i.e., it is a random 1-byte value bounded by the privacy ID pool. From $x_i$, one then generates a ciphertext: 
\[
y_i = E_{k}(iv, x_i)
\]
where $E$ denotes the AES/CBC/PKCS7 cipher, initialized with the key $k$ and the initialization vector $iv.$ 
The privacy ID corresponding to each $i$ is then computed by taking the first 12 bytes of $y_i$: 
\[ 
pid_i = y_i[0:11]. 
\]

\paragraph{BLE advertisement generation}
Finally, the advertisement payload is generated from the privacy ID combined with various meta data. Table \ref{fig:fmmadv} describes the full advertisement payload; Table \ref{tbl:fmmadv-support-info} provides details of the support info byte.

\definecolor{Gray}{gray}{0.85}
\newcolumntype{a}{>{\columncolor{Gray}}c}
\begin{table}[htp]
    \centering
    \caption{FMM lost mode advertisement structure}
    \label{fig:fmmadv}

    \resizebox{.8\columnwidth}{!}{%
    \begin{tabular}{|c|ccc|} 
    \hline
    Byte & 0  & 1-12  & 13 \\ 
    \hline
    Info & operation mode & privacy ID & support info \\ 
    \hline
    \end{tabular}
    }
\end{table}

\begin{table}[htp]
    \centering
    \caption{Support info byte (byte 13)}
    \label{tbl:fmmadv-support-info}
    \resizebox{.8\columnwidth}{!}{%
    \begin{tabular}{|c|ccccc|} 
    \hline
    Bit & 0-3 & 4 & 5 & 6 & 7 \\ 
    \hline
    Info & region info & E2E flag & UWB flag & MCF flag & reserved \\ 
    \hline
    \end{tabular}
    }
\end{table}
The first byte describes the operation mode of the OF protocol that is being used by a lost device. In the main OF protocol, this byte is always zero. The last byte contains information about the device’s region and functionalities supported. This last byte varies depending on the device but stays consistent for all advertisements for a device. If two lost mode devices are advertising in the same area, then this last byte can be used as a quasi-differentiator between the two, provided they do not have the same settings/support.

Once the advertising data has been generated, the lost device starts advertising over BLE on the OF service’s UUID FD69. The device will continuously advertise the same data until a timer is triggered that causes it to shuffle the advertising data. This timer is set to trigger every 60 minutes, after which the device generates a new random nonce to be used to generate the advertising data. Since, there are only 51 possible values for the nonce $i$, and it is the only source of non-determinism, there are also only 51 possible values of the advertising data (for a \texttt{PrivateIDConfig}). The lost device repeats this process until it is online again. If an adversary has access to the device’s secret key and IV, then it is trivial to generate the 51 possible values. 

\subsubsection{BLE operations for newer FMM versions}\label{sec:privID-obf}
\paragraph{Privacy ID obfuscation}
Samsung has introduced change to the lost mode advertisement format for Galaxy smart devices with FMM version 7.2.24.12 or above. 

\begin{table}[htp]
    \centering
    \caption{Payload format for new FMM lost mode advertising}
    \label{tab:adv-fmm-new}
    \resizebox{\columnwidth}{!}{%
    \begin{tabular}{|c|cccccc|} 
    \hline
    Byte & 0  & 1-12  & 13 & 14 & 15-17 & 18-19  \\ 
    \hline
     Info & operating mode & 
     (obfuscated) privacy ID & 
     support info & 
     reserved &
     aging counter &
     signature \\
    \hline
    \end{tabular}
    }
\end{table}

As shown in Table \ref{tab:adv-fmm-new}, the key differences for the new FMM advertisement structure include the following: 

\begin{description}
    \item[obfuscated privacy ID] an obfuscation method is applied to the privacy ID contained in the BLE data. Details of the obfuscation method will be explained next.
    \item[timestamp] the added timestamp field is used to store a 3-byte value that represents the time when the advertisement data was computed. It is computed by dividing the current system time by 900 and casting the result to an integer:  \[timestamp=currentTimeSeconds // 900\]
    \item[signature] the added signature field stores a 2-byte value. It is computed using \lstinline$HMAC-SHA256$ with the device's secret key, and first 18 bytes of the advertisement data.
\end{description}

It was observed that the new FMM version still uses a size-51 privacy ID pool as for the older versions, meaning that there will only use 51 unique raw privacy ID for each device. However, an obfuscation algorithm is used to add more randomness to the privacy ID value contained in the advertisement data.

The algorithm uses a universally static obfuscation table and deterministically generated random Bytes to obfuscate each raw privacy ID:

\begin{enumerate}
    \item First, the FMM app computes a 12-byte raw privacy ID using the same algorithm as in older versions
    
    \item Then, the app computes the current timestamp value and saves it to a variable, which is later used to determine the obfuscation filter being used and the random bytes generated by \lstinline$java.util.Random$.
    
    \item The obfuscation table is an \lstinline$ArrayList<String>$ object that contains four hex strings as shown in Table \ref{tbl:obfuscation-tbl}. For each advertisement data generation process, the FMM app selects the $timestamp \mod{4}$th string and convert it to a 12-byte filter.

    \begin{table}[htp]
        \centering
        \caption{Content of the obfuscation table}
        \label{tbl:obfuscation-tbl}
        \resizebox{.6\columnwidth}{!}{%
        \begin{tabular}{ll}
          \toprule
          Index & Value \\
          \midrule
            0 & 88DFAF0581FFCEB1429F2200 \\
            1 & 4A2635F7AD0E416906A35CBE \\
            2 & 19DB724B07DF72B9792511DE \\
            3 & 7BC79BAB386B8AFEFE63B9B7 \\
          \bottomrule
        \end{tabular}}
    \end{table}

    \item Then, the app creates a new instance of the \lstinline$Random$ class and uses the $timestamp$ as the seed to generate 12 random bytes deterministically.
    
    \item Finally, the obfuscated privacy ID is computed as follows:
    \[privacyID_{obf} = privacyID_{raw} \oplus filter \oplus randomBytes \]
\end{enumerate}

\paragraph{Privacy ID de-obfuscation}
The obfuscation operation can be easily reversed by an adversary to extract the raw privacy ID from the lost advertisement data observed over BLE:
 
    \begin{enumerate}
        \item The adversary uses the $timestamp$ value in the BLE data as the seed for \lstinline$java.util.Random$, which allows $randomBytes'$, the same random bytes used in the obfuscation process, to be generated.
        \item Then, the adversary selects $timestamp \mod{4}$th string from the universally static obfuscation table as the filter, $filter'$.
        \item Finally, the adversary can obtain the raw privacy ID from the following XOR operation:
        \[privacyID_{raw}=privacyID_{obf} \oplus filter' \oplus randomBytes'\]
    \end{enumerate}

\section{Further details of the SmartTag protocol}
\label{sec:smarttag-details}

\subsection{SmartTag GATT profile}

Figure~\ref{fig:gatt} shows the GATT profile for SmartTags, with some characteristics omitted, keeping only the ones relevant to this paper. 

\begin{figure*}[t]
{\tiny
\begin{tikzpicture}[align=center,>=latex',font=\small]
\tikzstyle{NormalGATT} = [draw=black,thick,rectangle,minimum width=160mm,minimum height=144mm]
\tikzstyle{NormalCHAR} = [draw=black,thick,rectangle,minimum width=72mm,minimum height=15mm]
\tikzstyle{NormalSERV} = [draw=black,thick,rectangle,minimum width=76mm,minimum height=57mm]
\node[label=above:GATT Profile,NormalGATT](GATTProfile){};

\node[label=above:Onboarding Service\\FD59, above right=6mm and 3mm of GATTProfile.south west,anchor=south west,NormalSERV](Onboarding){};
\node[below right=3mm and 2mm of Onboarding.north west,anchor=north west,NormalCHAR](MAC){IDENTIFIER (read)\\08A11E38-1C6D-4929-9C32-4F32A64985CE};
\node[below=18mm of MAC.north,anchor=north,NormalCHAR](HASH){HASHED\_SN (read)\\6AC16DB1-F442-4BF4-B804-04C32356465D};
\node[below=18mm of HASH.north,anchor=north,NormalCHAR](FD59CHARS){......};

\node[label=above:Authentication Service\\EEDD5E73-6AA8-4673-8219-398A489DA87C, below right=11mm and 3mm of GATTProfile.north west,anchor=north west,NormalSERV](AUTH){};
\node[below right=3mm and 2mm of AUTH.north west,anchor=north west,NormalCHAR](NONCE){NONCE (indicate, read, write)\\A12BE31C-5B38-4773-9B9D-3D5735233A7C};
\node[below=18mm of NONCE.north,anchor=north,NormalCHAR](ENONCE){ENONCE (indicate, write) \\4EBE81F6-B952-465E-9ECE-5CA39D4E8955};
\node[below=18mm of ENONCE.north,anchor=north,NormalCHAR](CIPHER){SUPPORTED\_CIPHER (read, write)\\50F98BFD-158C-4EFA-ADD4-0A70C2F5DF5D};

\node[label=above:DFU Service\\FE59, below left=11mm and 3mm of GATTProfile.north east,anchor=north east,NormalSERV](DFU){};
\node[below right=3mm and 2mm of DFU.north west,anchor=north west,NormalCHAR,minimum height=50mm](BUTTONLESSDFU){BUTTONLESS\_DFU (indicate, write)\\8EC90003-F315-4F60-9FB8-838830DAEA50};

\node[label=above:Command Service\\FD5A, above left=6mm and 3mm of GATTProfile.south east,anchor=south east,NormalSERV](Command){};
\node[below right=3mm and 2mm of Command.north west,anchor=north west,NormalCHAR](ALARM){OWNER\_ALARM\\(indicate, read, write)\\DEE30001-182D-5496-B1AD-14F216324184};
\node[below=18mm of ALARM.north,anchor=north,NormalCHAR](FIRMWARE){FIRMWARE \\(indicate, read, write no response)\\DEE3000C-182D-5496-B1AD-14F216324184};
\node[below=18mm of FIRMWARE.north,anchor=north,NormalCHAR](FD5ACHARS){......};
\end{tikzpicture}
}
\caption{The GATT Server profile for SmartTags}
\label{fig:gatt}
\end{figure*}

    \begin{table}[t]
        \centering
        \caption{Characteristics for exchanging public parameters}
        \label{tbl:public-param-chars}
        \resizebox{\columnwidth}{!}{%
        \begin{tabular}{lll}
          \toprule
          Name & UUID & Value \\
          \midrule
            CLOUD_PUBLIC_KEY & b5754629-6821-44c6-a118-492feecf6bb2 & 32-byte, varies \\
            RANDOM_VALUE & 6ac16db1-f442-4bf4-b804-04c32356465d & 32-byte, varies \\
          \bottomrule
        \end{tabular}}
    \end{table}
    
\begin{table*}[t]
    \centering
    \caption{Characteristics read for making the \lstinline$finalization request$}
    \label{tbl:finalization-char}
    {\small
    \resizebox{\textwidth}{!}{%
    \begin{tabular}{llll}
      \toprule
       Name & UUID & Value & Encrypted \\
      \midrule
        SPEC_VERSION/specVersion & dee3000e-182d-5496-b1ad-14f216324184 & \textcolor{blue}{``0.5.3"} & No \\
        FIRMWARE_VERSION/version & 30c48d2a-6ccb-4240-9f97-7f97a3f1c030 & \textcolor{blue}{``01.01.26"} & No \\
        MODEL_NAME/modelName & d19ddd83bbe14144bb18f3ceb57c480a & \textcolor{blue}{``EI-T5300"}  & No\\
        SUPPORTED_CIPHER/cipher  & 5b5f7a4c-257e-4841-92d5-0042658122b6 & \textcolor{blue}{``AES_128-CBC-PKCS7Padding"} & No \\
        CONFIGURATION_VERSION/configurationVersion & 12761292-241c-490c-8424-6f7cc8a8a027 & \textcolor{blue}{``2.0"} & Yes \\
        MNMN/mnmn  & 04052818-d201-43eb-9d81-e936dc86ee06 & \textcolor{blue}{``Samsung Electronics"} & Yes \\
        VID/vid & 77b08bec-5890-49d1-b021-811741b417e6 & \textcolor{blue}{``IM-SmartTag-BLE"} & Yes \\
      \bottomrule
    \end{tabular}
    }}
\end{table*}
    
        \begin{table*}[t]
        \centering
        \caption{Characteristics written to set up a tag for OF}
        \label{tbl:of-chars}
            {\small
            \begin{tabular}{llll}
              \toprule
               Name & UUID & Value & Encrypted\\
              \midrule
                REGION & bebfaa51-dcb8-44de-a4b8-fc8c9c7ef46d & a valid region code, e.g. 12 for AU & Yes \\ 
               Privacy ID Seed & d19ddd83-bbe1-4144-bb18-f3ceb57c480a & 12-byte, varies & Yes \\
               Privacy Pool Size & 7534c394-1f40-4d12-afd7-dc2a75bd6a44 & 1000 & Yes\\
               privacy ID IV & abd6e6ba-3843-4786-b9b2-b69548eed881 & 16-byte, varies & Yes\\
              \bottomrule
            \end{tabular}}
        \end{table*}

        \begin{table*}[t]
        \centering
        \caption{The \lstinline$SETUP_COMPLETE$ characteristic}
        \label{tbl:setup-complete-char}
            {\small
            \begin{tabular}{llll}
              \toprule
               Name & UUID & value & Encrypted\\
              \midrule
             SETUP\_COMPLETE & bcc8cce6-8af6-48dc-a0ae-547f7c095229 & ``FINISH" & Yes \\
              \bottomrule
            \end{tabular}}
        \end{table*}
        
    \begin{table*}[t]
        \centering
        \caption{The \lstinline$TIME_SYNC$ characteristic}
        \label{tbl:time-sync-char}
        {\small
        \begin{tabular}{llll}
          \toprule
           Name & UUID & Value & Encrypted \\
          \midrule
            TIME\_SYNC & dee30005-182d-5496-b1ad-14f216324184 & Will be set to the current UTC Time & Yes\\
          \bottomrule
        \end{tabular}}
    \end{table*}
    
    \begin{table}[htp]
        \centering
        \resizebox{\columnwidth}{!}{%
        \begin{tabular}{llll}
          \toprule
           Name & UUID & Value & Encrypted \\
          \midrule
            CONFIRM_STATUS & f299f805-17b3-43c1-ac12-fbcc59ee2f0d & 0x01 & Yes\\
          \bottomrule
        \end{tabular}}
    \end{table}

\subsection{SmartTag registration}\label{ssec:apx-smarttag-registration}

Here is a more concrete description of the registration process for a SmartTag. 

\subsubsection{Shared secret establishment}

\begin{description}[style=unboxed,leftmargin=0.5cm]
    \item[request (steps 1-2)]
    The phone makes a POST request to the \lstinline$/ide\-ntity/easysetup/blob$ URL of the \url{client smartthings.com} domain. 
    
    \begin{lstlisting}[frame=none,style=base,basicstyle=\scriptsize,numbers=none,linewidth=\columnwidth]
    {
        "keyid": {"type":"hashed_sn","value": @hashed_sn@},
        "mnId":"0AFD",
        "rand": @x@,
        "setupId":"430"
    }
    \end{lstlisting}
    
    The listing above shows the body of the \lstinline$shared secret request$, which consists of the following information collected from the tag's GATT characteristics and BLE advertisements:
    \begin{description}
        \item[general information] The values of \lstinline$mnId$ and \lstinline$setupId$ were obtained from the BLE advertisement data of the tag. 
        \item[information unique for each tag] The \lstinline$hashed_sn$ is the base64 encoding of tge first 6 bytes of the hashed serial number read from the \lstinline$HASHED_SERIAL_NUMBER$ characteristic. The random number \lstinline$x$ is a 32-byte value randomly generated by the phone for each registration session. Those two parameters are the public parameters for the shared secret establishment process.
    \end{description}
    
    \item[response (steps 3-4)]
    The server returns an HTTPS response containing the shared secret $B_{secret}$ and the public key $B_{pub}$ to the phone: after receiving the request, the server will find the public key of the tag $A_{pub}$ associated with the \lstinline$hashed_sn$. Then, the server will generate an ephemeral private-public key pair $b$ and $B_{pub}=bG$, and computes the ECDH shared key. Finally, the shared key is concatenated with the random number \lstinline$x$ to form the input for the SHA-256 hash function to produce the shared secret: $B_{secret}=SHA256(B_{key}|x)$.

    \begin{lstlisting}[frame=none,style=base,basicstyle=\scriptsize,numbers=none,label={fig:apx-verification-response}]
    {  "code":2000000,
       "message":"SUCCESS",
       "data":[
          {
             "type":"ECDH_ED25519",
             !"blob":"***",!
             "cloudPubKey":"***",
             "sn":"***",
             "meta":{
                "mfgr":"Samsung Electronics",
                "mnId":"0AFD",
                "onboardingId":"430",
                "modelName":"EI-T5300",
                "sku":[
                   "EI-T5300"
                ],
                ...
             }}]}
    \end{lstlisting}
   The Listing above shows the format of the server's response to a \lstinline$shared secret request$. The \lstinline$blob$ value corresponds to the shared secret.
\end{description}

\textbf{Steps 5-6} After receiving the response, the phone sends the ephemeral public key $B_{pub}$ and the random value $x$ to the tag via Write Requests to the corresponding GATT characteristics (see Table~\ref{tbl:public-param-chars}), allowing the tag to compute the shared secret in the same manner. 

\subsubsection{Ownership status check}

\begin{description}[style=unboxed,leftmargin=0.5cm]
    \item[request] 
    The owner's device makes a GET request to\\
    the \lstinline$/chaser/trackers/lostmessage$ URL of\\ the \url{client.smartthings.com} domain with the \lstinline{serialNumber}, \lstinline{modelName}, \lstinline{mnId}, and \lstinline{setupId} GET parameters provided, to check the ownership status of the tag. 
 
    \item[response]
     The Status code of the response is generally either 200 or 404:
     \begin{itemize}
         \item 404 means the profile does not exist. This should be the response for a non-registered SmartTag, as the tag's parameters should not correspond to any existing profile on the location server. 
         \item 200 means a profile that matches the parameters exists. A 200 response body should contain an \lstinline$own : Boolean$ field. \lstinline$true$ means the tag is currently registered to the requester. \lstinline$false$ means the tag is registered to another user.
     \end{itemize}
    
    Thus, the registration process would only proceed if the response status code is 404, or 200 and \lstinline$own==true$.

\end{description}

\subsubsection{Finalizing tag registration}

\begin{table*}[htp]
  \centering
    \caption{Format of a non-registered SmartTag's advertisement}
   \label{tbl:smarttag-ads}
    \begin{tabular}{lll}
      \toprule
      Byte(s) & Description & Value \\
      \midrule
        0 &  & 01  \\
        1-4 & manufacture ID (\texttt{mnId}) &  30414644 (0AFD) \\
        5-7 & setup ID (\texttt{setupId}) & 343330 (430) \\
        8-10 &  & 010501 \\
        11-13 & last two bytes of the MAC address & varies, e.g., 33444431 for 3D:D1 \\
      \bottomrule
    \end{tabular}
\end{table*}

\begin{description}[style=unboxed,leftmargin=0.5cm]
    \item[request]
     The owner's device makes a POST request to the \lstinline$/miniat\-ure/mobile$ URL of the \url{client.smartthings.com} domain. 
    
        \begin{lstlisting}[frame=none,style=base,basicstyle=\scriptsize,numbers=none,linewidth=\columnwidth]
        {
            "tagData":{
                "firmware":{
                        "specVersion":`"0.5.3"`,
                        "version":`"01.01.26"`
                    },
                    "mnId":"0AFD",
                    "modelName":`"EI-T5300"`,
                    "serialNumber": @sn@,
                    "setupId":`"430"`
                },
            "cipher": `"AES_128-CBC-PKCS7Padding"`,
            "configurationVersion": `"2.0"`,
            "identifier": @sn@,
            "deviceName": ...,
            "encryptionKey": @shared_secret@,
            "locationId": ...,
            "mnmn": `"Samsung Electronics"`,
            "roomId": ...,
            "vid": `"IM-SmartTag-BLE"`
        }
        \end{lstlisting}
        
        The listing above shows the body of the \lstinline$finalization request$, which consists of the following information:
        \begin{description}
            \item[general information] The values of \lstinline$mnId$ and \lstinline$setupId$ were obtained from BLE advertisement data as shown in Table \ref{tbl:smarttag-ads}. The values colored in blue were read from the corresponding GATT characteristics under the onboarding service (UUID FD59) (see Table~\ref{tbl:finalization-char} for details).
            \item[requester specified] The \lstinline$deviceName$ value is ``SmartTag" by default. It can be any custom value specified by the owner's device and determines the name of the registered device displayed in the SmartThings application. The \lstinline$locationId$ and \lstinline$roomId$ values are keys for users to access their registered devices stored in a specific room of a location. Each new account has a default location and two default rooms, each associated with a unique and static id. Devices registered to an account is accessed in a \lstinline$locationId -> roomId -> devices$ way.
            \item[unique for each tag] The values of \lstinline$serialNumber$ and\\ \lstinline$i\-dentifier$ equate the identity MAC address of the tag. The \lstinline$encryptionKey$ field contains the shared secret received from the earlier stage (\S\ref{ssec:shared-secret-establishment}).
        
        \end{description}

    \item[response]
    The server's response to this request contains configuration data associated with the tag shown in Listing \ref{lst:finalization-res}. 
    
    \begin{lstlisting}[frame=none,style=base,basicstyle=\scriptsize,numbers=none,linewidth=\columnwidth,caption={The \lstinline$finalization response$ body},label={lst:finalization-res}]
    {
        "deviceId":...,
        "metadata":
        {
            "regionCode":...,
            "privacyIdPoolSize":...,
            "privacyIdSeed":...,
            "privacyIdInitialVector":...,
            ...
        }
    }
    \end{lstlisting}

    The \lstinline$deviceId$ value is used to access the profile of the tag for various operations, e.g., removing the tag, pulling the location history of a lost tag. 

\end{description}

The four values contained in the metadata are sent to the tag via Write Requests to the corresponding GATT characteristics for the tag to generate BLE data for OF after completing the registration process (see Table~\ref{tbl:of-chars} for details). Finally, the phone will perform time synchronization with the tag through the \lstinline$TIME_SYNC$ characteristic, then write a value to the \texttt{SETUP\_COMPLETE} characteristic to indicate the completion of the registration process (see Table~\ref{tbl:time-sync-char} and Table~\ref{tbl:setup-complete-char}). The tag will then drop the GATT connection with the owner device to operate as a registered tag and broadcast OF data on a Non-Resolvable RPA.

\subsection{Owner-Tag GATT interaction}
We describe some interesting commands used in the owner and tag interactions that we have identified.

\paragraph{Playing sound on SmartTags}
The GATT characteristic \texttt{OWNER\_ALARM} (UUID DEE30001-182D-5496-B1AD-14F216324184) is used to configure the alarm on a tag. A phone can make a SmartTag ring by sending a command packet,  with the command opcode 0x01, through this characteristic. The opcode for turning off the alarm is 0x00. 

\paragraph{SmartTag ringing the owner device}\label{para:remote-ring}
The Remote Ring feature allows a tag to make its owner device ring by pressing the tag button. This feature is not enabled by default and needs to be explicitly enabled by the owner through the SmartThings app. The owner device can receive this command through subscribing to the indication of the \texttt{remoteRing} GATT characteristic (UUID DEE30003-182D-5496-B1AD-14F216324184) of the tag.

A tag's button has three states, which are \texttt{pushed}, \texttt{held}, and \texttt{pushed\_2x}. The Remote Ring is triggered by double-pressing the tag button. After the first press, the Button Characteristic will send an indication with the value 0x01, which corresponds to \texttt{pushed}. After the second press, the characteristic will send another indication with the value 0x03, which corresponds to \texttt{pushed\_2x}.

When the owner device receives 2 Button Characteristic indications, where the second indication has a value of 0x03 and a greater counter value than the first indication, the remote ring will be triggered.

\paragraph{Over-the-air firmware update}
 The vendor has implemented a custom channel for performing firmware updates on SmartTags. Over-the-air firmware updates can be performed by writing encrypted commands to the \lstinline$FIRMWARE$ characteristic (UUID DEE3000C-182D-5496-B1AD-14F216324184). After establishing an authenticated session. The SmartThings application will automatically read the firmware version of a tag from characteristic UUID DEE3000B-182D-5496-B1AD-14F216324184. If the firmware version is below the latest version, the application will ask the user to start a firmware update. Once the firmware update is triggered, the SmartThings app would download the latest firmware through an API call, which would generate a temporary link (that would expire after a couple of hours) to download the firmware (from the domain \url{smart-tag-firmware.samsungiotcloud.com}). The downloaded firmware will be cached locally. 
 
 
 To initiate a firmware update on the tag, the phone must first subscribe to the \lstinline{FIRMWARE} characteristic's indication, then execute the \texttt{transferFirmwareInformation} command by writing opcode 0x00 and a list of arguments to the characteristic, as shown in Table \ref{tbl:fwinfo}. The value of the \texttt{transferWindow} argument is 0x0A (10) in the BLE packets we observed, which means that the tag will send a handle value indication after receiving every ten firmware data packets and wait for the phone to confirm.

\begin{table}[htp]
  \centering
    \caption{Format of the transferFirmwareInformation command}
  \label{tbl:fwinfo}
    \begin{tabular}{*{3}{cll}}
      \toprule
      Opcode & Argument & Type\\
      \midrule
      \multirow{5}{*}{0x00} 
                & totalFirmwareSize & uint32 \\
                &  totalFirmwareCRC16 & uint16 \\
                &  newFirmwareVersionLength &  uint8 \\
                &  newFirmwareVersion & string \\
                & transferWindow & uint8\\
      \bottomrule
    \end{tabular}
\end{table}

 By analyzing the device logs produced during the update process, we could observe all the firmware update commands and arguments being executed. Table \ref{tbl:transferFirmwareInformationargs} shows the arguments for the \lstinline$transferFirmwareInfo\-rmation$ command, which indicates that the firmware has a total size of 179620 (0x2bda4), total CRC of 37858 (0x93e2), firmware version 1.02.06 (0x312e30322e3036), and transfer window size of 10 (0x0a).

\begin{table}[htp]
    \centering
    \caption{Captured transferFirmwareInformation command}
    \label{tbl:transferFirmwareInformationargs}
     \resizebox{.8\columnwidth}{!}{%
    \begin{tabular}{|c|c|c|c|c|c|c|} 
    \hline
    Byte & 0 & 1-4 & 5-6  & 7 & 8-14 & 15        \\ 
    \hline
    Value & 00 & a4bd0200 & e293 & 07 & 312e30322e3036 & 0a \\
    \hline
    \end{tabular}}
\end{table}

Then the phone will start to execute the \texttt{transferFirmwareData} command by writing opcode 0x01, followed by a list of arguments including segmented firmware data to the characteristic until all the firmware data is transferred. Table \ref{tbl:fwdata} shows the format of the \texttt{transferFirmwareData} command. 

\begin{table}[htp]
  \centering
    \caption{Format of the transferFirmwareData command}
  \label{tbl:fwdata}
    \begin{tabular}{*{3}{cll}}
      \toprule
      Opcode & Argument & Type\\
      \midrule
      \multirow{4}{*}{0x01} 
                & offset & uint32 \\
                &  segmentedFirmwareDataLength & uint16 \\
                &  segmentedFirmwareData &  byteArray \\
                &  argumentsCRC16 & uint8 \\
      \bottomrule
    \end{tabular}
\end{table}

\subsection{SmartTag removal}\label{sec:remove-a-tag}
A Registered SmartTag can be removed by its owner through SmartThings. If the target tag is connected to the Owner Device, the Owner Device will first perform a factory reset of the tag through GATT interaction by writing the \texttt{reset} command (opcode 0x01) to the \texttt{tag.factoryReset} characteristic (UUID dee30006-182d-5496-b1ad-14f216324184) under the Command Service. 

The Owner Device will also make an DELETE request to the \texttt{/devices/[deviceId]} URL of the \url{api.smartthings.com} domain to remove the online profile of the SmartTag. After the online profile of a tag is removed, GET request to the \url{/chaser/trackers/lostmessage?serialNumber=...&modelName=...&mnId=...&setupId=...} URL would receive a 404 Not Found error response from the server, indicating that no existing profile associated with the device defined by the serial number, manufacture ID, and setup ID is found.

\begin{lstlisting}[frame=none,style=base,basicstyle=\scriptsize,numbers=none,caption={SmartTag removal request}, label={lst:device-removal}]
DELETE /devices/!a976012b-****-****-****-25ba6379eff0! HTTP/1.1
X-ST-Client-OS: Android 8.0.0
X-ST-Client-DeviceModel: samsung SM-G935F universal8890
X-ST-Client-AppVersion: 1.7.89.25
User-Agent: ***
Authorization: Bearer ***
Accept: application/vnd.smartthings+json;v=1
Accept-Language: en-AU
X-ST-CORRELATION: 850fafd0-0527-4f13-864e-c64f9c7eb077
Host: api.smartthings.com
Connection: close
Accept-Encoding: gzip, deflate
\end{lstlisting}

\subsection{Location querying and reporting}\label{ssec:apx-smarttag-location}

\subsubsection{SmartTag BLE advertisement data}
This section provides further details on the BLE advertisement behavior of registered SmartTags (see \S\ref{para:smarttag-ble-adv}).

The tables below contains BLE data from a registered tag captured through BLE passive scanning.

It can be observed that the {\em operating state} changed from 5 (Connected to one device) to 1 (Premature Lost) immediately after the owner device disconnected. After another 15 minutes, {\em operating state} changed to 2 (Lost), and the tag would update {\em aging counter}, privacy ID, and the signature every 15 minutes.

\begin{table}[htp]
\centering
\begin{adjustbox}{width=\columnwidth}
\csvreader[
    tabular=*{10}{c},
    table head=\toprule,
    table foot=\bottomrule,
    column count=3,
    no head,
    late after line=\\,
    late after first line=\\\midrule]{test1.csv}{}{\csvlinetotablerow}
\end{adjustbox}
\end{table}

At 23:28 on 30 Aug, the {\em operating state} changed to 3 (Overmature Lost) after operating in state 2 for approximately 24 hours. 

\begin{table}[htp]
\centering
\begin{adjustbox}{width=\columnwidth}
\csvreader[
    tabular=*{10}{c},
    table head=\toprule,
    table foot=\bottomrule,
    column count=3,
    no head,
    late after line=\\,
    late after first line=\\, ]{test2.csv}{}{\csvlinetotablerow}
\end{adjustbox}
\end{table}

Data collected over the next 24 hours shows that the {\em aging counter} and signature would still update every 15 minutes in state 3, yet the updating frequency for the privacy ID reduced to once every 24 hours.

\begin{table}[htp]
\centering
\begin{adjustbox}{width=\columnwidth}
\csvreader[
    tabular=*{10}{c},
    table head=\toprule,
    table foot=\bottomrule,
    column count=3,
    no head,
    late after line=\\,
    late after first line=\\, ]{test3.csv}{}{\csvlinetotablerow}
\end{adjustbox}
\end{table}

\subsubsection{Location reporting}\label{sssec:apx-smarttag-location-reporting}
Location reports requires interactions with the \lstinline$Location Server$. This section details the protocol steps for obtaining a new JWE access token from the \lstinline$Location Server$.

\paragraph{Nonce request}
The client makes a GET request to the \lstinline$/nonce$ URL of the \lstinline$Location Server$ to obtain a 16-byte nonce randomly generated by the server, as shown in the example response:

\begin{lstlisting}[frame=none,style=base,basicstyle=\scriptsize,numbers=none, caption={Nonce response}]
HTTP/1.1 200 
Date: ...
Content-Type: application/json
Content-Length: 44
Connection: close

{"nonce":"0b4c****************************"}
\end{lstlisting}

\paragraph{Access token request}
Then, the client makes a POST request to the \lstinline$/accesstoken$ URL of the \lstinline$Location Server$ to obtain a new JWE access token.

\begin{lstlisting}[frame=none,style=base,basicstyle=\scriptsize,numbers=none, caption={Access token request}]
POST /accesstoken HTTP/1.1
x-iot-findnode-version: 722500014
!signature: UYkRnBsu1K...!
!certificate: MIID1zCCA...!
x-iot-findnode-type: MOVING
x-iot-findnode-host: GALAXY_PHONE
!nonce: 0b4c****************************!
Content-Type: application/json; charset=utf-8
User-Agent: ...
Host: chaser-ap03-apnortheast2.samsungiotcloud.com
Connection: close
Accept-Encoding: gzip, deflate
Content-Length: 2

{}
\end{lstlisting}

As shown in the above request, the request headers contain a \lstinline$certificate$, \lstinline$signature$, and a \lstinline$nonce$ field.

\begin{description}
    \item[certificate] The value of this field is produced by concatenating the two certificates contained in a native library file \lstinline$libfmm_ct.so$.
    \item[signature] The value of this field is produced by encrypting the SHA-256 hash of the \lstinline$nonce$ with the private key using the RSA algorithm. The private key is also loaded from the native library.
    \item[nonce] the nonce used to generate the signature
\end{description}

By analyzing the runtime memory of the FMM app, we were able to recover the value of the private key. 
For ethical concerns, we do not provide the private signing key here. However, we provide below here a sample signature, generated from the text \texttt{"Usenix Security 2023"} (without the quotes) using the private key,  the public key to verify the signature, and the certificate chain that certifies the authenticity of the public key. 

{\scriptsize 
\begin{verbatim}
text message: Usenix Security 2023

signature: 
UR1dEuaPaBZ87CqvBnLr/Y82uBA1vcrisHaeDujvYguqL1OAIkGnsYqs+Z6rNhC+
wta+Ri1p4QKyxFJtUxEFw6dp3+yJ0hTnFdhs8QjXv+AHKhQ0DQI8Bdpx0XYXMAzv
hkmDVOWwdGER69VAS2bztJZ0/I3Q+Hlnd4+KbVfgcN1npZovLGqQxXZA0FrKZKzL
HSi7YCmj8YJSigFE6MktH6xueb12efuMHxvXubDb3LXCNPMCACqCef224Z9sHbq0
myL7S+OyoNEP07uC2aNSABeDJfuj3gG6xOoMDGN72oRS87w9KEIsCDdhhQLnWkmh
AzHVQgcU7jWnxrjbJnjxEA==

Public key to verify the signature: 
-----BEGIN PUBLIC KEY-----
MIIBIjANBgkqhkiG9w0BAQEFAAOCAQ8AMIIBCgKCAQEA3hrhxMrVHd43kghgdlMJ
oDqIWqozPkq+GsGB9URKsFRBfhQud3tD3d5eenvw9OrlDHgxQiaCebPSt86Ff3Rx
sAvyp90G3eF1AyRj0MMech9LUY/rixoAFRe1FkJYuEVn8bkWN+EX0D8kRxR4Cft7
80CXfbTldt6jVGuAvM0LJOyeg9YTHR4DBFVQTozqX/C2LpwDtsRLkBrnfnMnCFyx
Pn/z8k9vcoM3Dtc5/WzU8z3z1VGblEmvmYvnh8d6Wogi23As6+/oBy875R846FIP
yzbElJJvmHc8jPPsMiolF9Hm2oyBkRR8SL7Z0E2i1GWyA3xccm6IPqEOPatwkJV5
QQIDAQAB
-----END PUBLIC KEY-----
\end{verbatim}
}

The X.509 certificate chain to certify the above public key is given below (in PEM format). The chain is incomplete as it is missing the root CA certificate, which we did not manage to find. 

{\scriptsize 
\begin{verbatim}
-----BEGIN CERTIFICATE-----
MIID1zCCAzmgAwIBAgIEWfnphzAKBggqhkjOPQQDAjBZMQswCQYDVQQGEwJLUjET
MBEGA1UEBxMKU3V3b24gY2l0eTEXMBUGA1UECxMOU2Ftc3VuZyBNb2JpbGUxHDAa
BgNVBAMTE1NhbXN1bmcgY29ycG9yYXRpb24wHhcNMTcxMTAxMTUzNDMxWhcNMzcx
MDI3MTUzNDMxWjCCAUMxCzAJBgNVBAYTAktSMQ0wCwYDVQQHDARHVU1JMSQwIgYD
VQQKDBtTYW1zdW5nIEVsZWN0cm9uaWNzIENvLiBMdGQxJzAlBgNVBAsMHk1vYmls
ZSBDb21tdW5pY2F0aW9ucyBEaXZpc2lvbjE1MDMGA1UEAwwsREN2VlhVbFZlTkRn
elN5S2R6MjNxYjd6UE1nN2FWNHlnODBFZGt0Qzdycz0xGDAWBgoJkiaJk/IsZAEZ
FghTTS1HOTY1RjGBhDCBgQYKCZImiZPyLGQBAQxzU0FLX1YxOjIwMTcxMTAyMDAz
NDI2OjEwNDpQcGtJUDF3Y3U1b0VfbmRNQkNNUkFSc052WTM0ZmhmdlFITnVLaXhM
RFRRPTpEQ3ZWWFVsVmVORGd6U3lLZHoyM3FiN3pQTWc3YVY0eWc4MEVka3RDN3Jz
PTB2MBAGByqGSM49AgEGBSuBBAAiA2IABFS0h2Utw+fTDOcVfj0pIoDkCV1QEAYB
pzS/Km1LVZYIbW5larGIbj6lTjRy06Z1RdQVN2rNuP6ve0WaX0oFAZy4PqOshJpq
NJToqwrlt7Wv8cwtkoHv/GVMlw+b/GE36KOB5TCB4jASBgNVHRMBAf8ECDAGAQH/
AgEAMA4GA1UdDwEB/wQEAwIBBjAdBgNVHQ4EFgQUf9rWjnMw3RgJOCAtDmTDKVMz
xkkwHwYDVR0jBBgwFoAUDaDvIFcFZ1y2gVUUSR33PJD0E7owPQYIKwYBBQUHAQEE
MTAvMC0GCCsGAQUFBzABhiFodHRwOi8vb2NzcC5zYW1zdW5nLmNvbS9zZWN1cml0
eS8wPQYDVR0fBDYwNDAyoDCgLoYsaHR0cDovL2NybC5zYW1zdW5nLmNvbS9zZWN1
cml0eS9yZGV2aWNlcy5jcmwwCgYIKoZIzj0EAwIDgYsAMIGHAkIBgS+Dl0dTfHC+
Yoi2tJ4N59r2hxeQA1jJal2T0OJjLFe5irz8U5Q8ZaPbu3xUc9PrCZLGBe+LmUf7
6T9/hNuCxnoCQViyQZ8C+1nLKejavgMmw1gx61uoqBg/JMWIWWkimXpain6NaHVN
NCg9ej0VshaY1CMHqKpSumiAs90mJTt+0Mft
-----END CERTIFICATE-----
-----BEGIN CERTIFICATE-----
MIIEyzCCBFCgAwIBAgIBATAKBggqhkjOPQQDAjCCAUMxCzAJBgNVBAYTAktSMQ0w
CwYDVQQHDARHVU1JMSQwIgYDVQQKDBtTYW1zdW5nIEVsZWN0cm9uaWNzIENvLiBM
dGQxJzAlBgNVBAsMHk1vYmlsZSBDb21tdW5pY2F0aW9ucyBEaXZpc2lvbjE1MDMG
A1UEAwwsREN2VlhVbFZlTkRnelN5S2R6MjNxYjd6UE1nN2FWNHlnODBFZGt0Qzdy
cz0xGDAWBgoJkiaJk/IsZAEZFghTTS1HOTY1RjGBhDCBgQYKCZImiZPyLGQBAQxz
U0FLX1YxOjIwMTcxMTAyMDAzNDI2OjEwNDpQcGtJUDF3Y3U1b0VfbmRNQkNNUkFS
c052WTM0ZmhmdlFITnVLaXhMRFRRPTpEQ3ZWWFVsVmVORGd6U3lLZHoyM3FiN3pQ
TWc3YVY0eWc4MEVka3RDN3JzPTAeFw0xNzExMDExNTM0MzFaFw0zNzEwMjcxNTM0
MzFaMB8xHTAbBgNVBAMMFEFuZHJvaWQgS2V5c3RvcmUgS2V5MIIBIjANBgkqhkiG
9w0BAQEFAAOCAQ8AMIIBCgKCAQEA3hrhxMrVHd43kghgdlMJoDqIWqozPkq+GsGB
9URKsFRBfhQud3tD3d5eenvw9OrlDHgxQiaCebPSt86Ff3RxsAvyp90G3eF1AyRj
0MMech9LUY/rixoAFRe1FkJYuEVn8bkWN+EX0D8kRxR4Cft780CXfbTldt6jVGuA
vM0LJOyeg9YTHR4DBFVQTozqX/C2LpwDtsRLkBrnfnMnCFyxPn/z8k9vcoM3Dtc5
/WzU8z3z1VGblEmvmYvnh8d6Wogi23As6+/oBy875R846FIPyzbElJJvmHc8jPPs
MiolF9Hm2oyBkRR8SL7Z0E2i1GWyA3xccm6IPqEOPatwkJV5QQIDAQABo4IBijCC
AYYwHwYDVR0jBBgwFoAUf9rWjnMw3RgJOCAtDmTDKVMzxkkwggEHBgorBgEEAdZ5
AgERBIH4MIH1AgEBCgEBAgECCgEBBBdjb20uc2Ftc3VuZy5hbmRyb2lkLmZtbQQA
MCy/hT0IAgYBZ3gf5Pu/hUUcBBowGDEUMBIEDUFuZHJvaWRTeXN0ZW0CAQExADCB
naEOMQwCAQACAQECAQICAQOiAwIBAaMEAgIIAKUXMRUCAQACAQECAQICAQMCAQQC
AQUCAQamETEPAgEBAgECAgEDAgEEAgEFv4FIBQIDAQABv4N3AgUAv4U+AwIBAr+F
QCowKAQg0cU7epMZCew38ZObFGIcbk/Rm/kHnRlfhrPOpHzR+S0BAf8KAQC/hUEF
AgMBX5C/hUIFAgMDFFMwLAYLKwYBBAGBbAsDFwcEHTAboBkTF2NvbS5zYW1zdW5n
LmFuZHJvaWQuZm1tMAsGA1UdDwQEAwIEsDAdBgNVHQ4EFgQUifL1UI1kDX9ibbF/
fPCXhGCYt/AwCgYIKoZIzj0EAwIDaQAwZgIxANgxyNmoG3E4K+vMmjjsc2VZyIEC
CqZas576Oa0U9IQYLY8zqlMImAM2VXcVeu17lgIxALL77S6iJJZ/C6ZshEmwV+nv
szt65vDjxQcrFXGGWdeSS8CByEAFowaaS2XruZIEPQ==
-----END CERTIFICATE-----
\end{verbatim}
}

The server's response for a valid request would contain a new authentication token that stays valid for 32 hours.

\begin{lstlisting}[frame=none,style=base,basicstyle=\scriptsize,numbers=none,linewidth=\columnwidth, caption={Access token response}]
HTTP/1.1 200 
Date: Sun, 16 Oct 2022 15:08:57 GMT
Content-Type: application/json
Content-Length: 893
Connection: close

{
   !"accessToken":"eyJhbGciOiJBMTI..."!,
   "expirationTime":166600...,
   "findNode":{
      "type":"MOVING",
      "host":"GALAXY_PHONE",
      "version":"...",
      "configuration":{
         ...
      }
   }
}
\end{lstlisting}

\paragraph{Location reporting request}
After obtaining a valid access token, a device can proceed to make location reports of lost SmartTags to the \lstinline$Location Server$. The Listing below shows an example location report, which contains the estimated geolocation of the lost tag and the BLE advertisement data of the tag (\lstinline$serviceData$).

\begin{lstlisting}[frame=none,style=base,basicstyle=\scriptsize,numbers=none,label={fig:tag-onlineprofile}]
POST /geolocations HTTP/1.1
Authorization: Bearer ***
Content-Type: application/json; charset=utf-8
...
      {
         "geolocation":{
            "accuracy":"16.145",
            "battery":"FULL",
            !"latitude":"***",!
            !"longitude":"***",!
            "method":"wifi",
            "rssi":"-56",
            "speed":"0.0",
            "timeStamp":1658235656202,
            "valid":true
         },
         "tagAdvertisement":{
            !"serviceData":"***"!
         }
    }
...
\end{lstlisting}

\subsubsection{Location querying}\label{ssec:owner-server-of}
An Owner Device receives location updates of its lost SmartTag by querying the \lstinline$SmartThings Server$. Listing \ref{lst:location-history-req} shows an example of a location history request. 

\begin{lstlisting}[frame=none,style=base,basicstyle=\scriptsize,numbers=none,caption={Location history request}, label={lst:location-history-req}]
POST /installedapps/********-****-****-****-************/execute
...
{
   "parameters":{
      "requester":"***",
      "clientType":"aPlugin",
      "extraUri":!"/trackers/********-****-****-****-************/geolocations?
            order=asc&startTime=1662991200001&endTime=1663077599999&
            isSummary=true&limit=200"!,
      "method":"GET",
      "encodedHeaders":"***",
      "requesterToken":"***",
      "encodedBody":"",
      "clientVersion":"1",
      "uri":"/trackerapi"
   }
}
\end{lstlisting}
The \lstinline$extraUri$ field contains the query to be executed by the server. For the above example, the server would return a list of a maximum number of 200 location reports received between 1662991200001 and 1663077599999 (GMT) for the SmartTag. The server can return at most 200 locations in the response and location data older than a week cannot be returned.



\section{Further analysis details}\label{sec:analysis-details}

\subsection{OF device emulation}\label{ssec:apx-impersonation}
This section provides further details of the OF device impersonation technique discussed in Section~\ref{sec:impersonation}. 

The shared secret used in the \lstinline$finalization request$ and the configuration data returned in the server's response can be used to set up the impersonation script. An impersonated SmartTag advertises BLE data in any user-specified tag state, e.g., user can configure the impersonated tag to always advertise in lost mode. The impersonated tag can also updates its aging counter, privacy ID, and signature fields in the BLE data periodically. The impersonated tag uses an active GATT server with the same architecture as a real SmartTag and supports various interaction with its Owner Device over BLE, e.g., BLE authentication, GATT Command encryption/decryption.

\begin{figure*}[t]
   \includegraphics[width=\textwidth]{{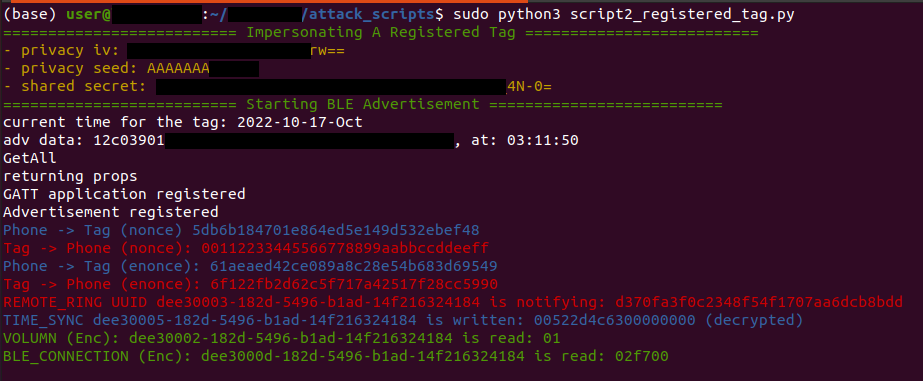}}
   \caption{Impersonated tag interacting with its owner}
   \label{fig:impersonation}
\end{figure*}

Figure \ref{fig:impersonation} shows the output produced by the impersonation script, which contains the configuration data for the impersonated tag, the BLE data it advertised, and the data exchange with its owner device over BLE. The logging corresponds to the following interactions: the impersonated tag advertises BLE data; completes BLE Authentication with its owner device; then executes the \lstinline$remoteRing$ command (see Appendix \ref{para:remote-ring}) to ring the owner device.

\subsection{Flaws in the SmartTag registration process}
\label{sssec:registration-attacks}

Recall that in the protocol for finalizing tag registration, the first step is the message from the owner device to the server:
\[
1. ~ O \to S ~ : ~ sn, id, B_{secret}
\]
In a normal registration flow, $sn$ is identical to $id$. However, we found that if we use a value $id$ that is different from $sn$, the registration would still be successful, and more importantly, the serial number contained in $id$ would then be marked as {\em registered} in Step 6 of the protocol. 
In other words, it would appear that the server does not check whether $sn = id$. Recall that $sn$ is the value used for checking ownership status (see \S\ref{ssec:validation}) of the tag. This can be exploited to register a tag multiple times: the attacker uses the serial number of an unregistered tag for the value of $sn$, and the serial number of an already registered tag for the value of $id.$ However, for the purpose of location querying and reporting, multiple registrations using the same serial number seem to be treated as different devices. 



\paragraph{Flaw 1. Tag data can be inferred without BLE communication}
data contained in the GATT characteristics of the tag is read by the owner device to communicate necessary information with the \lstinline$SmartThings Server$ for registration. This is essentially done by the two HTTPS requests:
    
\begin{enumerate}
    \item The \lstinline$shared secret request$ to obtain the shared secret from Samsung's server. 
    \item The \lstinline$finalization request$ to finalize the registration process and obtain configuration data required to set up the tag for operating in the registered mode. 
\end{enumerate}

Details of each request was discussed in Appendix \ref{ssec:apx-smarttag-registration}, and the data contained in each request was categorized into \textbf{general information}, \textbf{requester specified information}, and \textbf{information unique for each tag}. Most data exchanged during the BLE connection with the tag being registered is in the \textbf{general information} category.

Experiments have shown that the \textbf{general information} data can be treated as static in registration requests, and hence the hashed serial number \lstinline$hashed_sn$ and the serial number/identifier \lstinline$sn$ are the only two values unique for each tag. Since the \lstinline$hashed_sn$ is simply the SHA-256 hash of the serial number \lstinline$sn$, knowing a tag's identity address, which equates to \lstinline$sn$, allows an adversary to construct the HTTPS requests for registration without communicating with the tag over BLE connection.

\paragraph{Flaw 2. Lack of consistency check for the \lstinline$shared secret$}
The shared secret establishment protocol is based on the assumption that the requester of the \lstinline$shared secret request$ is an honest device. When an adversary gains the position of a requester, the registration process can be completed and a registered (impersonated) tag can be set up without computing the shared secret.

For instance, an adversary can complete the registration flow by directly communicating with the \lstinline$SmartThings Server$ via HTTPS requests, instead of using the SmartThings application (normal flow). The \lstinline$shared secret response$ would contain the shared secret and the \lstinline$finalization response$ would contain the configuration data (privacy ID configuration data) required to set up a registered tag through impersonation (details will be provided in later sections).

It was also discovered that the \lstinline$encryptionKey$ field in the finalization request can accept any custom 32-byte base64 encoded value to set up a device profile. Therefore, the shared secret establishment process can be skipped entirely when the registration requests are controlled by an adversary. More specifically, the adversary can skip the \lstinline$shared secret request$, and register a device directly by issuing the \lstinline$finalization request$ with a custom \lstinline$encryptionKey$ value as the shared secret for the device profile created by this request.

\paragraph{Flaw 3. Lack of consistency check for the device serial number}
Recall that the JSON body of the \lstinline$finalization request$ contains a \lstinline$identifier$ and a \lstinline$serialNumber$ field. In requests issued by honest devices, these fields always have the same value, which equals to the identity MAC address of the tag that is being registered. After receiving a \lstinline$finalization request$, the server would:

\begin{enumerate}
    \item compare the provided \lstinline$serialNumber$ with device serial numbers in its database.
    \item compare the provided \lstinline$identifier$ with the device identifiers in its database and each stored identifier is linked to its owner's account. 
\end{enumerate}

Step 1 is to confirm that the \lstinline$serialNumber$ value matches the serial number of a manufactured device, while step 2 is to confirm that the \lstinline$identifier$ value does not conflict with any existing device profile of another user. Hence, when trying to register a tag that is currently owned by others, the \lstinline$finalization request$ would fail with an ``Conflict" error. 

However, if the adversary that controls the HTTPS request changes the \lstinline$identifier$ value to a ``fake identifier" that differs from the identifier of the target tag, while the \lstinline$serialNumber$ value remains the same. The server was tricked into processing the request to created a new device profile for the adversary despite conflicting \lstinline$serialNumber$ value.

\paragraph{Attack 1}
By exploiting flaw 1 and 2, an adversary can register others' non-registered SmartTags, without having physical access to the tags when the identity MAC addresses of the tags are known to the adversary. The attack uses the identity MAC address \lstinline$sn$ of the target tag. The attack flow can be summarized as follows:
\begin{enumerate}
    \item  make an \lstinline$ownership status request$ to check whether the tag is non-registered (proceed to the next step if the tag is non-registered)
    \item  make the \lstinline$finalization request$ with a custom \lstinline$encryptionKey$ to register the device
    \item make an \lstinline$owner location report$ using the \lstinline$device_id$ provided in the \lstinline$finalization response$ to create an OF device profile on the \lstinline$Location Server$
\end{enumerate}
After the above steps, the ownership status of the tag should be linked to the SmartThings account for making the adversary's requests. Therefore, re-sending the \lstinline$ownership status request$ would receive: 
\begin{lstlisting}[frame=none,style=base,basicstyle=\scriptsize,numbers=none,linewidth=\columnwidth]
{'own': True, 'lostMessage': {'type': 'PREDEFINED'}}
\end{lstlisting}

\paragraph{Attack 2}
Attack 2 is an extension of Attack 1 by exploiting flaw 3 in addition, which allows an adversary to register a tag even when it is already owned by another user. The attack uses the identity address \lstinline$sn$ of the target tag. The attack flow can be summarized as follows:
\begin{enumerate}
    \item  make an \lstinline$ownership status request$ to check whether the ownership status. For a tag registered to another user, the response would be:
    \begin{lstlisting}[frame=none,style=base,basicstyle=\scriptsize,numbers=none,linewidth=\columnwidth]
    {'own': False, 'lostMessage': {'type': 'PREDEFINED'}}
    \end{lstlisting}
    
    \item To bypass the duplication check, we proceed make the \lstinline$finalization request$ with \lstinline$identifier=sn$, and \\ \lstinline$serialNumber=fake_sn$, where \lstinline$fake_sn$ has a different value from  \lstinline$sn$, it can be a randomly generated 6-byte hex string.
\end{enumerate}

The server response to the \lstinline$finalization request$ indicates the success of bypassing the duplication check, logging into the SmartThings app (using the adversary's account) after the above steps will see the new tag in the devices tab. 

\paragraph{Implications}
The above attacks allow attackers to achieve the following goals: registering a non-registered SmartTag and preventing the owner of the actual tag from using it (Attack 1); registering a SmartTag that is already owned by another user (Attack 2). Users of any SmartThings software/SmartTag firmware versions can potentially be affected by the attacks. 

Both attacks requires the adversary to know the identity address of the target tag, which can be done in several ways:
\begin{itemize}
    \item non-registered tags: a non-registered tag advertises BLE data on its identity address, making it trivial for a nearby adversary to obtain
    \item registered tags: a registered tag advertises data on a Non-Resolvable RPA, yet there are multiple ways to obtain its identity address:
    \begin{enumerate}
        \item adversary with physical access to a registered tag can factory reset the tag physically to obtain its identity address.
        \item adversary without physical access, but within BLE communication range with a registered tag can exploit the DFU vulnerability to infer its identity address from the DFU MAC address (see \S\ref{ssec:dfu}).
    \end{enumerate}
\end{itemize}

Additionally, Attack 2 may also be used to create multiple independent device profiles using a single valid identity address, where each device profile can be set up manually using the impersonation script (see \S\ref{ssec:apx-impersonation}). This essentially allows an adversary to use the FMM service for free, as the adversary can set up an arbitrary number of tags through the attack and impersonation without the need of possessing any SmartTags.

\paragraph{Mitigation}
The above vulnerabilities may be mitigated as follows:
\begin{itemize}
    \item For Flaw 1: When reverse-engineering the SmartTag's firmware, we noticed that a SmartTag seems to have a static EdDSA private key embedded in the hardware (just like the ECDH private key). 
    
    Therefore, the registration protocol can be extended by a signature verification step to ensure BLE communication with the target tag: the server sends a random nonce to the requester for each registration request; to finalize the registration, the requester must return the signature produced by signing the nonce with the EdDSA private key of the tag; The server then verifies this signature using the EdDSA public key associated with the tag and only finalizes the registration process if the verification succeeds.
    
    
    \item For Flaw 2: a consistency check mechanism can be added to prevent any protocol step of the registration flow being skipped, e.g., an adversary should not be able to skip the \lstinline$shared secret request$ and make a \lstinline$finalization request$ directly, and to ensure that values presented at different protocol steps are consistent, e.g., the \lstinline$encryptionKey$ in the \lstinline$finalization request$ cannot differ from the shared secret returned by the server in the \lstinline$shared secret response$.
    
    \item For Flaw 3: a consistency check mechanism can also be added to ensure the values within the same request are consistent, e.g., \lstinline$serialNumber=identifier$ for the \lstinline$serialNumber$ and the \lstinline$identifier$ values in the \lstinline$finalization request$.
    
\end{itemize}

\begin{figure*}[t]
  \centering
  \includegraphics[width=\textwidth]{{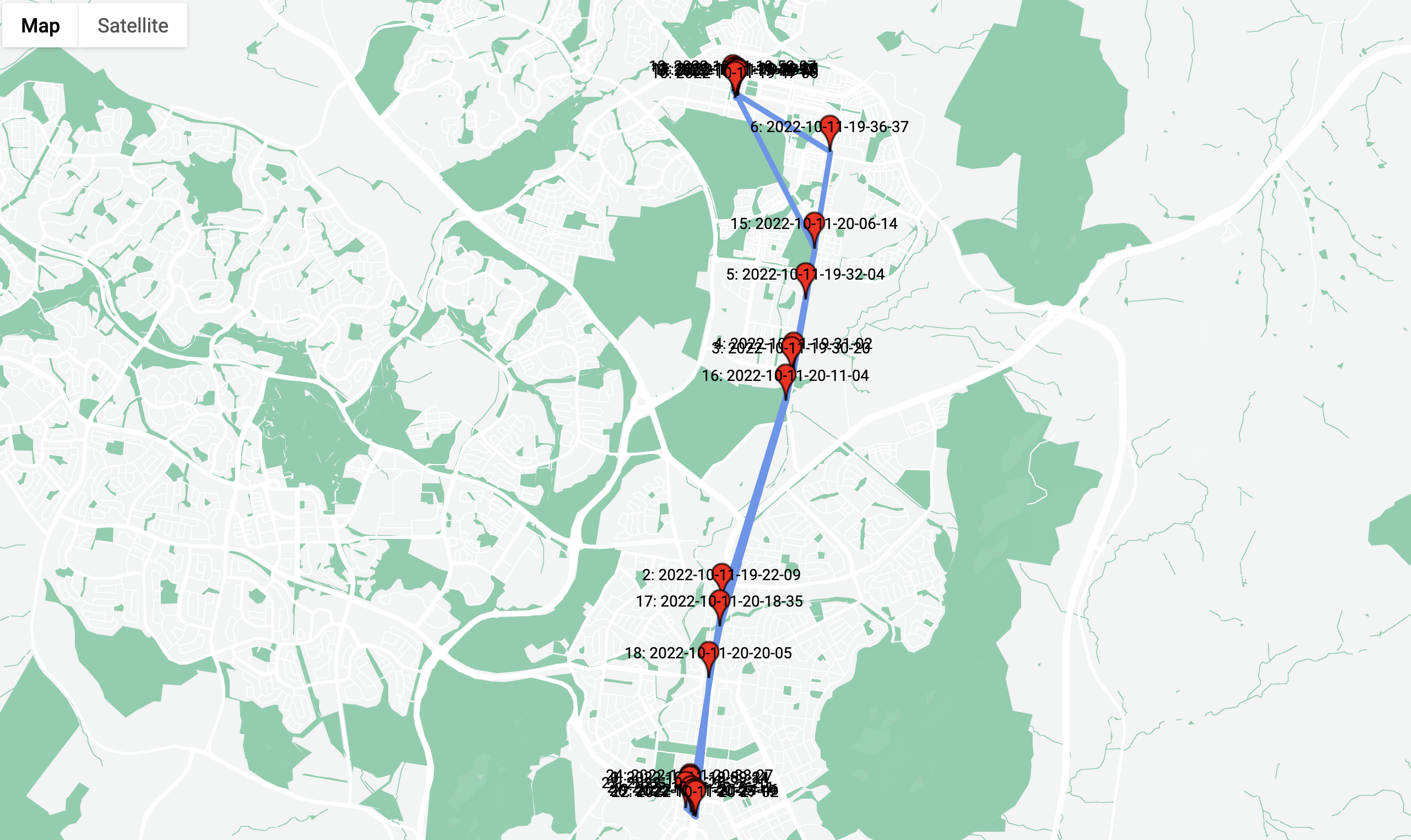}}
  \caption{Estimated path plotted from raw locations provided by Samsung's server}
  \label{fig:tracking-experiment}
\end{figure*}

\subsection{Unwanted tracking}\label{apx-unwanted-tracking}
This section extends our discussion in \S\ref{sec:unwanted-tracking} by demonstrating the risks of SmartTag tracking through an experiment. 

During the experiment, the lost mode tag we carried for approximately one and a half hour (from 18:59:24 to 20:33:27 on October 11th, 2022) was reported 25 times by nearby helper devices. Then, we plotted the estimated path travelled by the tag based on the geolocations and their timestamps queried from the Samsung server using the owner's account. The plotted path is shown in Figure \ref{fig:tracking-experiment}. 

We show that location reports submitted by Helper Devices in Samsung's OF network can provide precise location information of a lost tag, increasing the potential consequences of malicious tracking attacks, as an attacker can receive real-time updates of the tracker tag's location and hence infer the routine and location of the victim.

\end{document}